\pdfoutput=1

\documentclass{ws-ijmpd}
\usepackage[super,compress]{cite}
\usepackage{hyperref}
\hypersetup{pdftitle = The title of my PDF, pdfauthor = My name, pdfsubject= The subject, pdfkeywords = keyword1 keyword2 keyword3}
\hypersetup{colorlinks = true, linkcolor = blue, anchorcolor = red, citecolor = blue, filecolor = red, pagecolor = red, urlcolor = blue}

\begin{document}

\markboth{E.E. Zotos \& F.L. Dubeibe}
{Orbital dynamics of the Sun-Jupiter system}

%
\catchline{}{}{}{}{}
%

\title{ORBITAL DYNAMICS IN THE POST-NEWTONIAN PLANAR CIRCULAR RESTRICTED SUN-JUPITER SYSTEM}

\author{EUAGGELOS E. ZOTOS}

\address{Department of Physics, School of Science, \\
Aristotle University of Thessaloniki, \\
GR-541 24, Thessaloniki, \\
Greece\\
evzotos@physics.auth.gr}

\author{F. L. DUBEIBE}

\address{Facultad de Ciencias Humanas y de la Educaci\'{o}n, \\
Universidad de los Llanos, Villavicencio, Colombia \\
Grupo de Investigaci\'{o}n en Relatividad y Gravitaci\'{o}n, \\
Escuela de F\'{i}sica, Universidad Industrial de Santander, \\
A.A. 678, Bucaramanga 680002, \\
Colombia\\
fldubeibem@unal.edu.co}

\maketitle

\begin{history}
\received{29 September 2017}
\revised{3 November 2017}
\accepted{14 November 2017}
\published{5 December 2017}
\end{history}

\begin{abstract}
The theory of the post-Newtonian (PN) planar circular restricted three-body problem is used for numerically investigating the orbital dynamics of a test particle (e.g., a comet, asteroid, meteor or spacecraft) in the planar Sun-Jupiter system with a scattering region around Jupiter. For determining the orbital properties of the test particle, we classify large sets of initial conditions of orbits for several values of the Jacobi constant in all possible Hill region configurations. The initial conditions are classified into three main categories: (i) bounded, (ii) escaping and (iii) collisional. Using the smaller alignment index chaos indicator (SALI), we further classify bounded orbits into regular, sticky or chaotic. In order to get a spherical view of the dynamics of the system, the grids of the initial conditions of the orbits are defined on different types of two-dimensional planes. We locate the different types of basins and we also relate them with the corresponding spatial distributions of the escape and collision time. Our thorough analysis exposes the high complexity of the orbital dynamics and exhibits an appreciable difference between the final states of the orbits in the classical and PN approaches. Furthermore, our numerical results reveal a strong dependence of the properties of the considered basins with the Jacobi constant, along with a remarkable presence of fractal basin boundaries. Our outcomes are compared with earlier ones, regarding other planetary systems.
\end{abstract}


\ccode{PACS Number(s): 04.25.Nx, 05.45.-a, 45.50.Pk, 45.50.Jf}

\section{Introduction}
\label{intro}

In the strong field regime, the space-time describing binary systems is strongly nonlinear and time-dependent due to the emission of gravitational waves. The complexity of this scenario is one of the reasons to explain why no exact solution to Einstein's equations modeling a binary system in orbital motion has ever been found \cite{PW14}. Several highly sophisticated approaches to tackle this problem have been developed, including numerical solutions and approximation methods. In some systems with small orbital velocities (compared to the speed of light) and weak gravitational fields, the flagship tool is the post-Newtonian (PN) approximation \cite{W11}, developed since the youth of general relativity \cite{EIH38}. A natural scenario to observe non-negligible PN effects over long time-scales is the Solar System. Indeed, it is a well-known fact that the PN corrections must  be accounted for in cases such as in calculating the perihelion precession of Mercury's orbit.

On the other hand, since Poincar\'{e}'s demonstration about the impossibility to find an exact solution for the motion of two planets around the Sun \cite{P92}, the so-called three-body problem, and the existence of heteroclinic intersections in celestial bodies dynamics \cite{A63}, it is generally accepted that the Solar System may be chaotic \cite{BL08} and therefore, there exist practical limits for the long-time predictions \cite{L96}. Taking into account that the Newtonian three-body problem is just a first approximation of a much more complex setting, it is desirable to refine the current understanding of the Solar System dynamics by including general relativistic corrections. To do so, some authors have derived the first-order PN equations of motion (1-PN) for the circular restricted three-body problem \cite{B72,B91,C76,K67,MD94}, by using the Einstein-Infeld-Hoffmann (EIH) theory \cite{EIH38}.

Some studies have been carried out on the existence and stability of the equilibrium points of the relativistic problem of three bodies. The triangular libration points were studied in Ref. \refcite{BH97}, finding that in contrast to the classical restricted three-body problem, in the relativistic system the equilibrium points $L_4$ and $L_5$ are unstable. Ref. \refcite{DP02} controverted this result, due to the existence of a region of linear stability in the parameter space $(\mu, 1/c^{2})$, with $\mu$ the mass parameter and $c$ the speed of light. A detailed analysis of the collinear points for several Sun-planet pairs was performed in Ref. \refcite{RPKV01}, finding that for all cases examined the collinear points $L_1$, $L_2$, and $L_3$ are unstable. Concerning the dynamics of the PN restricted three-body problem, recent studies have shown that the PN terms can be understood as non-negligible perturbations to the classical system \cite{DLGG17}, and hence, for small distances between the primaries the PN dynamics differ qualitatively from the Newtonian one \cite{HW14}.

Undoubtedly, one of the most important aspects of the restricted three-body problem (RTBP) is the classification of the initial conditions of the orbits. Knowing the nature of the orbits has numerous modern applications, such as in space flight missions, in launching and positioning artificial satellites, while it also serves the basis of several planetary and exoplanetary theories \cite{G01a,G01b,G01c,G01d,SYC08}. It all started with the pioneer works of Refs. \refcite{N04,N05}, where the first thorough and systematic orbit classification, in the classical RTBP, took place. During the last few years, the orbital dynamics of many planetary systems has been numerically investigated, through the classification of initial conditions in some particular scattering regions; the most notable works are the following: regarding the Earth-Moon system \cite{dAT14}, for the Saturn-Titan system \cite{Z15a} and for the Pluto-Charon system \cite{Z15b}.

In this paper, we shall explore the orbital dynamics of the Sun-Jupiter system, when the PN correction terms are included in the equations of motion as well as in the variational equations. The scattering region will be located in the vicinity around Jupiter and particularly between the Lagrange points $L_1$ and $L_2$. Our main goal is to classify initial conditions of the orbits of the test particle and determine the corresponding types of basins (bounded, escape and collision).

The paper is organized as follows: In section \ref{props} we first identify the most important properties of the dynamical system. Next, we outline the computational methods used for the classification of the initial conditions of the orbits in section \ref{cometh}. In the following section, a thorough and systematic numerical investigation takes place, thus revealing the orbital dynamics of the planar PN Sun-Jupiter system. Finally, our paper ends with section \ref{disc} where the discussion is given.

\section{Properties of the dynamical system}
\label{props}

\subsection{Description of the mathematical model}
\label{mod}

With some necessary simplifying assumptions, the orbital motion of Jupiter around the Sun can be considered circular (even though the eccentricity of its orbit is about 0.04839266). Under this condition, the motion of a test particle (e.g. a comet, asteroid or spacecraft) can be modeled as a planar circular restricted three-body problem (henceforth PCRTBP), in which the two bodies, the primary (Sun) $m_{1}$  and the secondary (Jupiter) $m_{2}$,  move in circular orbits, with the same angular velocity, around their common centre of mass, at a fixed distance $a$ \cite{MQ14}. The test particle $m_{3}$, whose mass is negligible compared to $m_{1}$ and $m_{2}$, does not perturb the motion of the primary and secondary bodies, and moves on the same plane $(x,y)$ under the combined gravitational influence of the two main bodies.

In order to facilitate the analysis of the system we use the Szebehely convention \cite{S67} for the normalization of the constants
$$m_1 = 1 - \mu,\quad m_2 = \mu,\quad x_1 = - \mu,\quad x_2 = 1 - \mu = 1 + x_1,$$
where $\mu = m_{2}/(m_{1} + m_{2}) \in [0, 1/2]$, is the mass parameter, where $m_1 > m_2$. With this choice of units, $a = 1$ and the origin O is located at the centre of mass of the two main bodies. Moreover, the centers $P_1$ and $P_2$ of the two main bodies are located at $(x_1, 0)$ and $(x_2, 0)$, respectively.

To account for general relativistic effects, we shall consider the first-order PN correction terms introduced by the Einstein-Infeld-Hoffmann (EIH) theory \cite{EIH38}. In a synodic frame of reference, the equations of motion for the test particle, in accordance with Ref. \refcite{MD94}, are given by the following set of differential equations
\begin{align}
\ddot{x} &= x + 2 \dot{y} - \frac{m_1\left(x - x_1\right)}{d_1^3} - \frac{m_2 \left(x - x_2\right)}{d_2^3} + \frac{\epsilon}{c^2} {\cal R}_{x}, \label{eq1}\\
\ddot{y} &= y - 2 \dot{x} - y \left(\frac{m_1}{d_1^3} + \frac{m_2}{d_2^3}\right) + \frac{\epsilon}{c^2} {\cal R}_{y}, \label{eq2}	
\end{align}
where $x_{i}$ denotes, in the synodic frame of reference, the fixed position of the body $m_{i}$, while $d_{i} = \sqrt{\left(x - x_{i}\right)^{2} + y^{2}}$. Furthermore, ${\cal R}_{x}$ and ${\cal R}_{y}$ are the relativistic correction terms.

The system of equations (\ref{eq1}-\ref{eq2}) admits one integral of motion, which is known as the Jacobi integral (or Jacobi constant) and can be written as \cite{MD94}
\begin{align}
J(x,y,\dot{x},\dot{y}) &= \frac{m_1}{d_1} + \frac{m_2}{d_2} + \frac{1}{2}\left(x^2 - \dot{x}^2 + y^2 - \dot{y}^2\right) \nonumber\\
&+ \frac{\epsilon}{c^2} {\cal J}_{R} = C,
\label{ham}
\end{align}
where $C$ is the Jacobi constant which is conserved.

The parameter $\epsilon$ is a transition parameter where $\epsilon = 0$ indicates classical Newtonian dynamics, while $\epsilon = 1$ corresponds to PN dynamics. In what follows, we shall only consider the PN case $(\epsilon = 1)$, unless the contrary is explicitly stated.

According to Ref. \refcite{DLGG17} the relativistic correction terms read
\begin{align}
{\cal R}_{x}&=
   \frac{m_1 m_2}{a}
   \left(\frac{x-x_1}{d_1^3}+\frac{x-x_2}{d_2^3}\right)
   +y \left(\frac{m_1}{d_1^3}+\frac{m_2}{d_2^3}\right)
   \nonumber\\
   &
   \times \left(\dot{x}-y\right)\left(x+4 \dot{y}\right)
   +\left(\frac{m_1
   \left(x-x_1\right)}{d_1^3}+\frac{m_2 \left(x-x_2\right)}{d_2^3}\right)
   \nonumber\\
   &
   \times \Bigg[4
   \left(\frac{m_1}{d_1}+\frac{m_2}{d_2}\right)
   -3\dot{x} \left(y-\dot{x}\right)
   -\left(x+\dot{y}\right)^2\Bigg]
   \nonumber\\
   &
   -\frac{3}{2} \left(\frac{m_1 \left(x-x_1\right)x_1^2}{d_1^3}
   +\frac{m_2 \left(x-x_2\right) x_2^2}{d_2^3}\right)
   \nonumber\\
   &
   +\frac{3}{2} y^2
   \left(\frac{m_1 \left(x-x_1\right) x_1^2}{d_1^5}+\frac{m_2
   \left(x-x_2\right) x_2^2}{d_2^5}\right)
   \nonumber\\
   &
   +2\omega_{1} \left(x+\dot{y}\right)
   +\left(\frac{7 x}{2}+4 \dot{y}\right)
   \nonumber\\
   &
   \times \left(\frac{m_1 \left(x-x_1\right) x_1}{d_1^3}
   +\frac{m_2 \left(x-x_2\right)x_2}{d_2^3}\right)
   \nonumber\\
   &
   -\frac{7}{2} \left(\frac{m_1 x_1}{d_1}+\frac{m_2
   x_2}{d_2}\right),
\label{ct1}
\end{align}
\begin{align}
{\cal R}_{y}&=
   \frac{m_1
   m_2}{a} y\left(\frac{1}{d_1^3}+\frac{1}{d_2^3}\right)
   +\frac{3}{2} y^3 \left(\frac{m_1
   x_1^2}{d_1^5}+\frac{m_2 x_2^2}{d_2^5}\right)
   \nonumber\\
   &
   +2\omega_{1} \left(y-\dot{x}\right)
   +\left(\frac{m_1}{d_1^3}+\frac{m_2}{d_2^3}\right)
   \nonumber\\
   &
   \times \Bigg\{
   y \left[4
   \left(\frac{m_1}{d_1}+\frac{m_2}{d_2}\right)
   +3 \dot{y}
   \left(x+\dot{y}\right)-y^2\right]
   \nonumber\\
   &
   +\dot{x} \left(3 x^2+3 x \dot{y}+2
   y^2\right)-\dot{x}^2 y
   \Bigg\}
   -\left[7 x
   \left(\dot{x}-\frac{y}{2}\right)+3 \dot{x} \dot{y}\right]
   \nonumber\\
   &
   \times \left(\frac{m_1
   x_1}{d_1^3}+\frac{m_2 x_2}{d_2^3}\right)+\left(4 \dot{x}-\frac{5
   y}{2}\right) \left(\frac{m_1 x_1^2}{d_1^3}+\frac{m_2 x_2^2}{d_2^3}\right)
   \nonumber\\
   &
   -\left(y-\dot{x}\right)
   \left(x+\dot{y}\right) \left(\frac{m_1 \left(x-x_1\right)}{d_1^3}+\frac{m_2
   \left(x-x_2\right)}{d_2^3}\right),
\label{ct2}
\end{align}
and
\begin{align}
{\cal J}_{R} &=
   \frac{1}{8}\left[\left(x^2+y^2\right)^2-3 \left(\dot{x}^2+\dot{y}^2\right)^2\right]-\frac{1}{4}\left(x^2+y^2\right)\left(\dot{x}^2+\dot{y}^2\right)
   \nonumber\\
   &
   -\left(x\dot{y}-y\dot{x}\right) \left(\dot{x}^2+\dot{y}^2\right) + \omega_{1} \left(x^2+y^2\right) -\frac{1}{2}\left(x\dot{y}-y\dot{x}\right)^2
   \nonumber\\
   &
   +\frac{3}{2} \left(\frac{m_{1}}{d_{1}}+\frac{m_{2}}{d_{2}}\right)\left(x^2 -\dot{x}^2+y^2-\dot{y}^2\right)
   \nonumber\\
   &
   +\frac{3}{2} \left(\frac{m_{1} x_{1}^2}{d_{1}}+\frac{m_{2} x_{2}^2}{d_{2}}\right)
   - \frac{m_{1} m_{2}}{a}\left(\frac{1}{d_{1}}+\frac{1}{d_{2}}\right)
   \nonumber\\
   &
   - \frac{y^2}{2} \left(\frac{m_{1} x_{1}^2}{d_{1}^3}+\frac{m_{2} x_{2}^2}{d_{2}^3}\right)
   -\frac{1}{2}  \left(\frac{m_{1}}{d_{1}}+\frac{m_{2}}{d_{2}}\right)^2
   \nonumber\\
   &
   -\frac{7}{2} x \left(\frac{m_{1} x_{1}}{d_{1}}+\frac{m_{2} x_{2}}{d_{2}}\right),
\label{ct3}
\end{align}

The angular velocity $\omega_{1}$ was calculated by \cite{C76} and is given by
\begin{equation}
\omega_{1} = \frac{m_{1}m_{2} - 3 (m_1 + m_2)^2}{2 a (m_1+m_2)},
\label{omega}
\end{equation}
which reduces to $\omega_{1}=[\mu(1-\mu)-3]/2$, with the help of the mass parameter.

It is important to note that in contrast to the classical Newtonian PCRTBP, the PN system depends on two parameters, the mass parameter $\mu$ and the speed of light $c$. According to Ref. \refcite{nasa16} for the Sun-Jupiter system the mass parameter is $\mu =$ 0.000953817733371, and the speed of light in canonical units is $c = 22945.236186$ (see \ref{Ap1}).

\subsection{Equilibrium points}
\label{pts}

The equilibrium points (or Lagrange points) correspond to equilibrium points of the PCRTBP, and can be found by replacing $\ddot{x} = \ddot{y} = \dot{x} = \dot{y} = 0$ in the respective equations of motion (\ref{eq1}-\ref{eq2}). In analogy with the classical Newtonian case, the PN PCRTBP has five equilibrium points (also known as Lagrange points): three of them, $L_1$, $L_2$, and $L_3$, are collinear points located on the $x$-axis, while the other two $L_4$ and $L_5$ are called triangular points and they are located on the vertices of an equilateral triangle. The central stationary point $L_1$ is located between the primary and the secondary, $L_2$ is at the right side of the secondary (Jupiter), while $L_3$ is at the left side of the primary (Sun).

We have solved numerically the resulting algebraic system $\left.\ddot{x}\right\vert_{\dot{x} = \dot{y} = 0} = 0$ and $\left.\ddot{y}\right\vert_{\dot{x} = \dot{y} = 0} = 0$ finding approximate expressions for the Lagrange points in terms of the mass parameter $\mu$. For the collinear points, we find that the classical expansion in powers of $n/3$ (see e.g. Ref. \refcite{F12}) along with an expansion of $\mu$ in powers of $n$, gives a very accurate solution with a maximum error of $10^{-14}$. Hence, the solution can be written as
\begin{equation}
x^{*}_{L_{i}} = \sum_{n=0}^{15}C_{n}^{i}\mu^{n}+\sum_{n=1}^{10}\tilde{C}_{n}^{i}\left(\frac{\mu}{1-\mu}\right)^{n/3},
\label{col}
\end{equation}
where $i=1,2,3$, denotes the $i$-th collinear point. In a similar fashion, for the triangular points we get
\begin{equation}
x_{L_j}^{*} = \sum_{n=0}^{2}T_{n}^{i}\mu^{n}, \quad y_{L_j}^{*}=\sum_{n=0}^{2}\tilde{T}_{n}^{i}\mu^{n},
\label{triag}
\end{equation}
where $j=4,5$, denotes the $j$-th triangular point. The numerical coefficients entering equations \eqref{col} and \eqref{triag} are given in \ref{Ap2}.

Additionally, in Table \ref{tab1} we present the numerical values of the Lagrange points for the Sun-Jupiter system in the framework of the PN-approximations (column 3) compared to the classical Newtonian values (column 2). It deserves mentioning that the numerical results presented in Table \ref{tab1} satisfy the respective system of algebraic equation, with an accuracy of $10^{-4}$ m, which indicates that the differences presented in column 4 are not numerical artifacts.

Our results suggest that the largest deviation is observed for the $x$-coordinate of $L_{4}$ and $L_5$, while the smallest deviation takes place for the $x$-coordinate of $L_{3}$. In general, the displacements of the Lagrange points are such that the new positions are closer to the primaries, i.e. $L_{1}, L_{2}, L_{4}$ and $L_{5}$ are moved toward Jupiter, while $L_{3}$ is shifted toward the Sun.

\begin{table*}
 \center
 \caption{Planar coordinates $(x,y)$ of the Lagrange points $L_{i}$ measured in metres, for the classical Newtonian and the PN PCRTBP. The last column denotes the corrections to the distances introduced by the PN terms.}
 \label{tab1}
 \begin{tabular}{lccr}
  \hline
  Coordinate & Newtonian Gravity $(m)$ & PN-Approximation $(m)$ & Difference $(m)$ \\
  \hline\hline
  $x(L_{1})$             &  7.257656518990008$\times10^{11}$ &  7.257656519293031$\times10^{11}$ &   30.302 \\
  $x(L_{2})$             &  8.319894593317031$\times10^{11}$ &  8.319894592936772$\times10^{11}$ &  -38.025 \\
  $x(L_{3})$             & -7.787213864029970$\times10^{11}$ & -7.787213864019399$\times10^{11}$ &    1.057 \\
  $x(L_{4}) =  x(L_{5})$ &  3.884635501925640$\times10^{11}$ &  3.884635511148704$\times10^{11}$ &  922.306 \\
  $y(L_{4}) = -y(L_{5})$ &  6.741245897986063$\times10^{11}$ &  6.741245892657048$\times10^{11}$ & -532.902 \\
  \hline
 \end{tabular}
\end{table*}

The values of the Jacobi integral at the Lagrange points $L_k$, $k = 1,..., 5$ are denoted by $C_k$ and they are critical levels with values: $C_1$ = 1.519379668835193, $C_2$ = 1.518743663753772, $C_3$ = 1.500476898588919, and $C_4 = C_5$ = 1.499523545304652.

\subsection{Hill region configurations}
\label{hrc}

\begin{figure*}[!t]
\centering
\resizebox{\hsize}{!}{\includegraphics{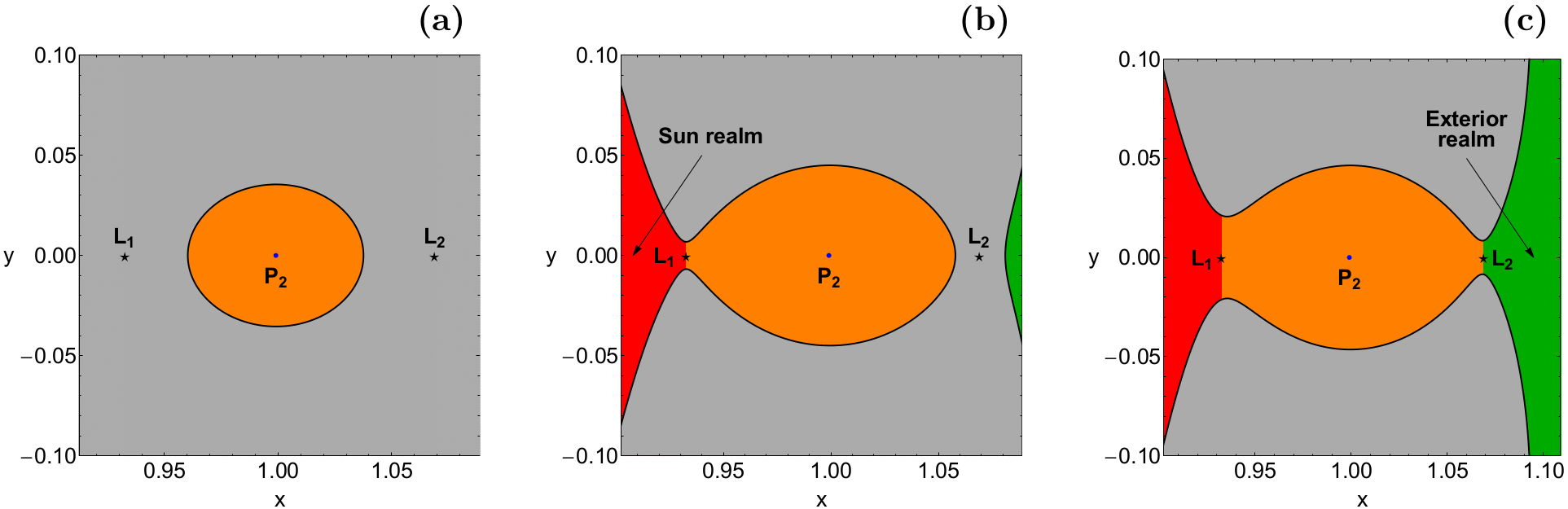}}
\caption{Characteristic examples of the first three Hill region configurations around the vicinity of Jupiter. The orange domains correspond to the scattering region of the Hill regions of the allowed motion, gray shaded domains indicate the energetically forbidden regions, while the thick black lines depict the Zero Velocity Curves (ZVCs). Furthermore the Sun and the exterior realms are indicated by red and green colors, respectively. The black stars pinpoint the position of the Lagrange points $L_1$ and $L_2$, while the position of the centre of Jupiter $(P_2)$ is indicated by a blue dot. (a): $C = 1.525$; (b): $C = 1.5193$; (c): $C = 1.51865$. (Colour online only.)}
\label{isos}
\end{figure*}

The projection of the four-dimensional phase space onto the configuration $(x,y)$ plane is called the Hill region configurations, which constitute the energetically accessible regions to the orbits for a given value $C$. The boundaries of these regions are called Zero Velocity Curves (ZVCs) because they are the locus in the configuration space where the kinetic energy vanishes. The value of the Jacobi constant strongly dictates the structure of the corresponding Hill region configurations. More precisely, there are five distinct cases:
\begin{itemize}
  \item Case I: $C > C_1$: All channels are closed, so there are only bounded and collisional motion (see panel (a) of Fig. \ref{isos}).
  \item Case II: $C_2 < C < C_1$: Only the channel around $L_1$ is open thus allowing the test particle to enter the Sun realm (see panel (b) of Fig. \ref{isos}).
  \item Case III: $C_3 < C < C_2$: The channel around $L_2$ opens, so the test particle can enter the exterior region and escape from the system (see panel (c) of Fig. \ref{isos}).
  \item Case IV: $C_4 < C < C_3$: Both channels, around the Lagrange points $L_2$ and $L_3$ are open, therefore the test particle is free to escape through two different directions.
  \item Case V: $C < C_4$: The energetically forbidden regions disappear, so motion over the entire configuration $(x,y)$ space is possible.
\end{itemize}
In Fig. \ref{isos}(a-c) we present the structure of the first three Hill region configurations.

\section{Computational methods and criteria}
\label{cometh}

For revealing the orbital dynamics in the PN Sun-Jupiter system, we need to numerically integrate the equations of motion (\ref{eq1}-\ref{eq2}), for several sets of initial conditions. For this purpose we consider dense uniform grids of $1024 \times 1024$ initial conditions $(x_0, y_0)$ regularly distributed on the configuration $(x,y)$ plane inside the energetically allowed area defined by the corresponding Jacobi constant $C$. Following a typical approach, all orbits are launched with initial conditions inside a certain scattering region which in our case is $x(L_1) \leq x \leq x(L_2)$ and $-0.1 \leq y \leq 0.1$. All orbits have $\dot{x_0} = 0$, while the initial value of $\dot{y}$ is always derived by the Jacobi integral (\ref{ham}) as $\dot{y_0} = \dot{y}(x_0,y_0,\dot{x_0},C) > 0$.

The classification of the initial conditions of the orbits is a rather demanding task if taking into account that the configuration space extends to infinity. In this study, orbits are classified, according to the types of motion, into three main categories:
\begin{itemize}
  \item Bounded orbits around the secondary (Jupiter).
  \item Orbits that escape from the scattering region.
  \item Orbits that collide with the secondary (Jupiter).
\end{itemize}
Moreover, all bounded orbits will be further classified into three sub-categories:
\begin{itemize}
  \item Non-escaping regular orbits.
  \item Trapped sticky orbits.
  \item Trapped chaotic orbits.
\end{itemize}

For distinguishing between regular and chaotic dynamics we use the Smaller Alignment Index (SALI) method \cite{S01} which has been proved to be a very fast yet reliable tool. The mathematical definition of SALI is the following
\begin{equation}
\rm SALI(t) \equiv min(d_-, d_+),
\label{sali}
\end{equation}
where $d_-$ and $d_+$ are the alignments indices. For computing the SALI we track simultaneously the time-evolution of the main orbit and the two deviation vectors. The nature of an orbit can be determined by the numerical value of SALI at the end of the numerical integration $t_{\rm max}$. In particular, if SALI $> 10^{-4}$ the orbit is said to be regular, whereas if SALI $< 10^{-8}$ we can thus conclude that the orbit is chaotic. On the other hand, when the value of SALI lies in the interval $[10^{-8}, 10^{-4}]$ we have the case of a sticky orbit\footnote{With the term ``sticky orbit" we refer to a special type of orbit which behaves as a regular one for long integration times before it exhibits its true chaotic nature.} and further numerical integration is needed to reveal the true character of the orbit.

In our numerical integrations the maximum time is 5000 dtu (dimensionless time units), corresponding to about 9445.93 Julian years. It should be mentioned, that orbits which do not escape or collide to Jupiter after a numerical integration of 5000 dtu are considered as bounded orbits (regular, sticky or chaotic).

Our next task is to define appropriate numerical criteria in order to distinguish between the above-mentioned types of motion. In order to consider a more realistic approach, we assume that Jupiter is a finite body, taking into account its mean radius approximately by 66854 km (about $8.58851 \times 10^{-5}$ dimensionless length units). The numerical integration stops when an orbit reaches the surface of Jupiter, thus producing an orbit leaking in the configuration space. Furthermore, an escaping orbit to the Sun realm must satisfy the condition $x < x(L_1) - \delta_1$, with $\delta_1 = 0.05$, whereas an escaping orbit to the exterior realm must fulfill the condition $x > x(L_2) + \delta_2$, with $\delta_2 = 0.04$. At this point, we must clarify that the tolerances $\delta_1$ and $\delta_2$ have been included in the escape criteria in order to avoid the incorrect classification of the unstable Lyapunov orbits \cite{L07} as escaping orbits.

The equations of motion (\ref{eq1}-\ref{eq2}) are numerically integrated by means of a double precision Bulirsch-Stoer integrator, using a numerical routine written in standard \verb!FORTRAN 77! \cite{PTVF92}, with a fixed time step equal to $10^{-2}$. Here it should be noted that the Bulirsch-Stoer integrator is both faster and more accurate than a double precision Runge-Kutta-Fehlberg 7(8) algorithm with Cash/Karp coefficients \cite{DMCG12}. Throughout all our computations, the Jacobi constant of Eq. (\ref{ham}) was conserved with fractional accuracy of about $10^{-11}$, or even better. The Lemaitre's global regularization method is applied in the case of collision orbits \cite{S67}, when the test particle moves around Jupiter into a region of radius $10^{-2}$.

For the numerical integration of the sets of initial conditions of orbits, in all types of two-dimensional colour-coded diagrams, that we will be presented in the following section, we needed roughly between 17 hours and 3.5 days of CPU time on an Intel$^{\circledR}$ Quad-Core i7 2.4 GHz PC. Moreover, all graphical illustrations presented in this paper have been created using the latest version 11.2 of the software Mathematica$^{\circledR}$ \cite{W03}.

\section{Orbit classification}
\label{clas}

In this section we will perform a thorough analysis of initial conditions of orbits in all possible Hill region configurations. Parallel to the classification we shall also record the time scale (or time period) of the collision and the time scale of the escapes.

In the following subsections we shall present colour-coded diagrams, thus following the methods also used in Refs. \refcite{Z15a,Z15b}. In these diagrams, each pixel is assigned a specific colour according to the particular type of the nature of the orbit. These colour-coded diagrams are, in a way, a modern version of the classical Poincar\'{e} Surface of section, where the phase space is a complex mixture of basins\footnote{By the term ``basin" we refer to a local set of initial conditions which lead to a certain final state (collision, escaping or bounded motion).} of bounded motion (regular or chaotic), escape and collision. Our numerical calculations indicate that the vast majority of bounded basins correspond to regular orbits, where a third integral of motion is present. This additional integral poses new restrictions to the available phase space and therefore it prevents them from escaping to infinity.

\subsection{Case I: $C > C_1$}
\label{ec1}

\begin{figure*}[!t]
\centering
\resizebox{\hsize}{!}{\includegraphics{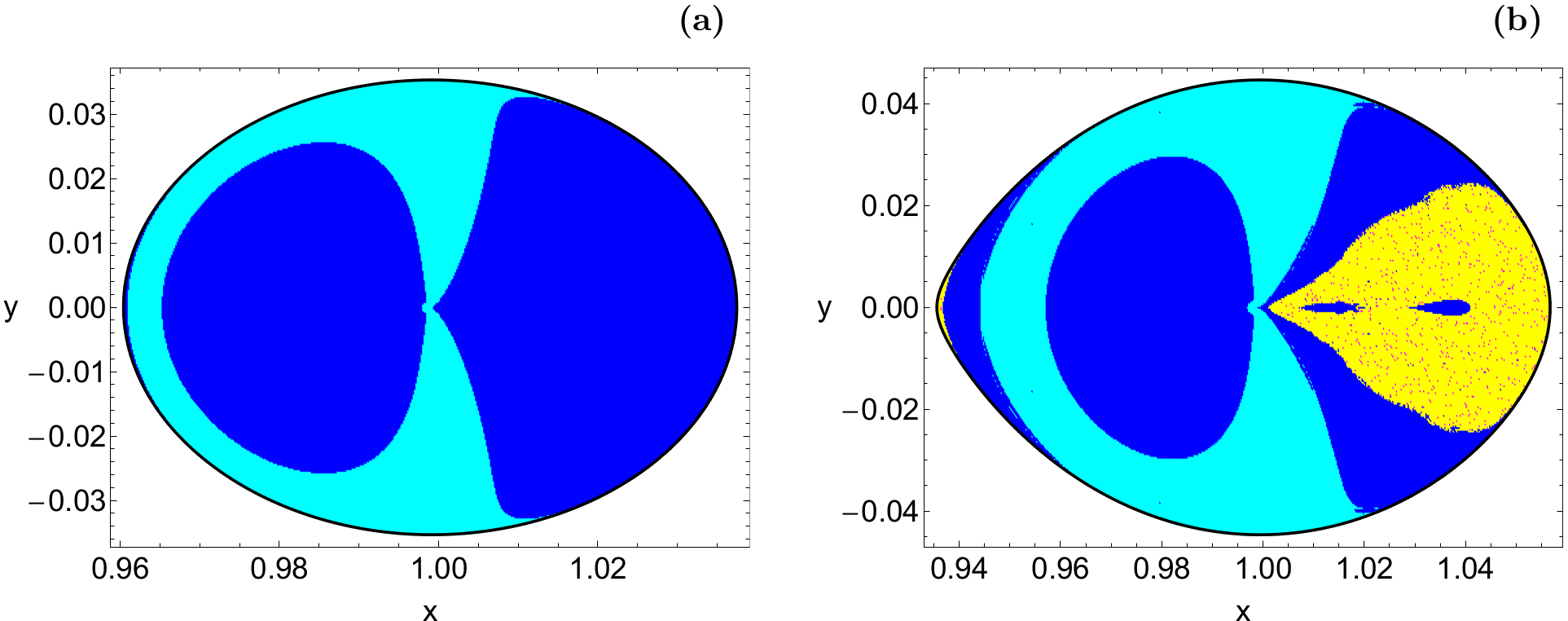}}
\caption{Colour-coded basin diagrams for Case I. (a-left): $C = 1.525$, (b-right): $C = C_1$. The colour code is as follows: non-escaping regular orbits (blue), trapped sticky orbits (magenta), trapped chaotic orbits (yellow), collision orbits (cyan). (Colour online only.)}
\label{hr1}
\end{figure*}

\begin{figure*}[!t]
\centering
\resizebox{\hsize}{!}{\includegraphics{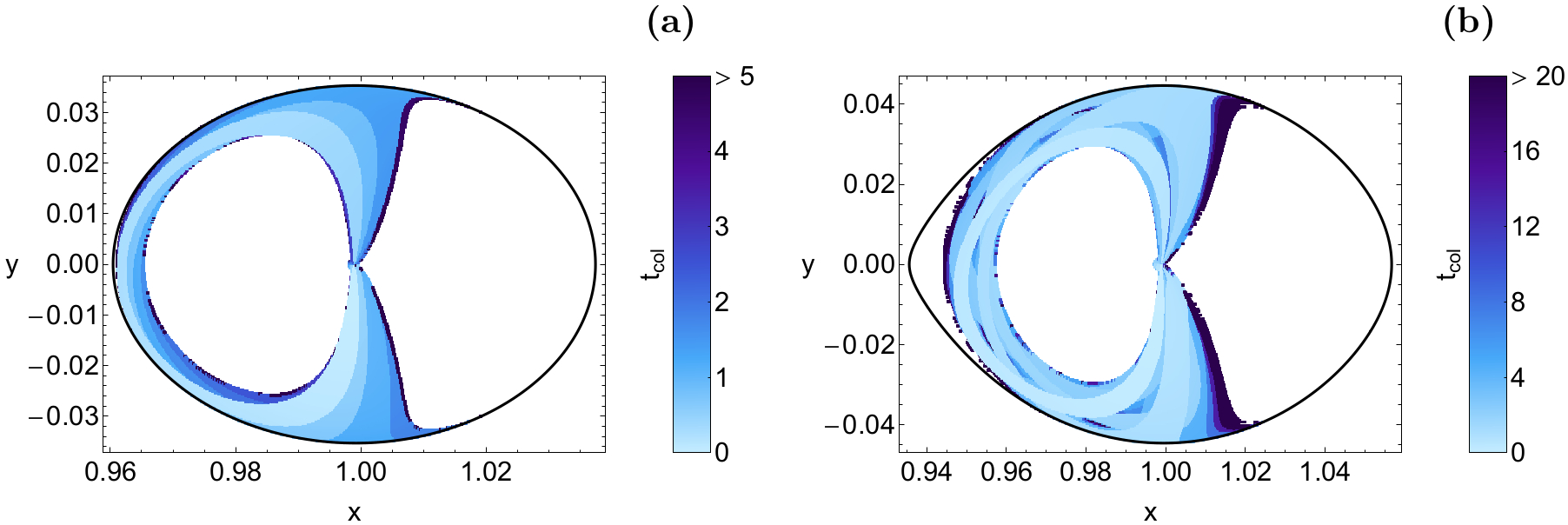}}
\caption{Distribution of the collision time of the orbits for the values of the Jacobi constant of Fig. \ref{hr1}(a-b). Large collision times are identified by darker colour, while the initial conditions of all types of bounded orbits are shown in white. (Colour online only.)}
\label{hr1t}
\end{figure*}

Our investigation begins with the scenario where the Hill region configurations consist of small disks around the secondary (Jupiter). In this case only two types of motion are possible: collision and bounded motion. The basin diagrams for two values of the Jacobi constant are presented in Fig. \ref{hr1}(a-b). The outermost black solid line is the ZVC which is defined as $J(x,y,\dot{x}=0,\dot{y}=0) = C$.

In panel (a) of Fig. \ref{hr1}, where $C = 1.525$, we see that two main stability islands of regular motion are present, surrounded by a unified collision basin. The bounded basin located on the left side of Jupiter contains initial conditions of retrograde (clockwise) quasi-periodic orbits around $P_2$. Such periodic orbits have a reflection symmetry, with respect to the horizontal $x$ axis. On the other hand, the stability island on the right side of Jupiter is composed of initial conditions that correspond to quasi-periodic orbits around $P_2$, travelling in counter-clockwise (prograde) sense, with respect to the rotating frame of reference. It is known \cite{SS00} that the regular orbits on the left side of Jupiter are much more stable than those on the right side of the secondary, in relation to the variation of the Jacobi constant $C$.

Panel (b) of Fig. \ref{hr1} illustrates the orbital structure of the configuration $(x,y)$ plane when $C = C_1$. It is seen that a third thin basin of initial conditions of non-escaping regular orbits emerges near the left boundary of the ZVC. There is no doubt that the main difference with respect to what we seen in panel (a) of Fig. \ref{hr1} concerns the area at the right side of Jupiter. We observe that the corresponding stability island which was present for $C = 1.525$ has split into pieces, while a trapped chaotic domain appears into the right side of Jupiter. With a much closer look we may identify, inside the trapped chaotic area, several isolated initial conditions which correspond to sticky orbits with sticky period larger than 5000 dtu. It should be emphasized that in Ref. \refcite{Z15b} similar results have been reported, regarding the Pluto-Charon system, for a Jacobi value very close to the critical value $C_1$. However in that case there was no further classification of the bounded orbits into sub-categories (ordered, sticky and chaotic).

In the following Fig. \ref{hr1t}(a-b), using tones of blue, we show the distribution of collision times on the configuration $(x,y)$ space. Light colors correspond to fast collision orbits, dark colors indicate large collision times, while white colour denote all types of bounded motion. It is interesting to note that in both energy levels the initial conditions of collision orbits form complicated colour layers (zones). Furthermore, one may observe that the orbits with the highest values of collision times have initial conditions mainly located in the vicinity of the boundaries of bounded basins.

\subsection{Case II: $C_1 > C > C_2$}
\label{ec2}

\begin{figure*}[!t]
\centering
\resizebox{\hsize}{!}{\includegraphics{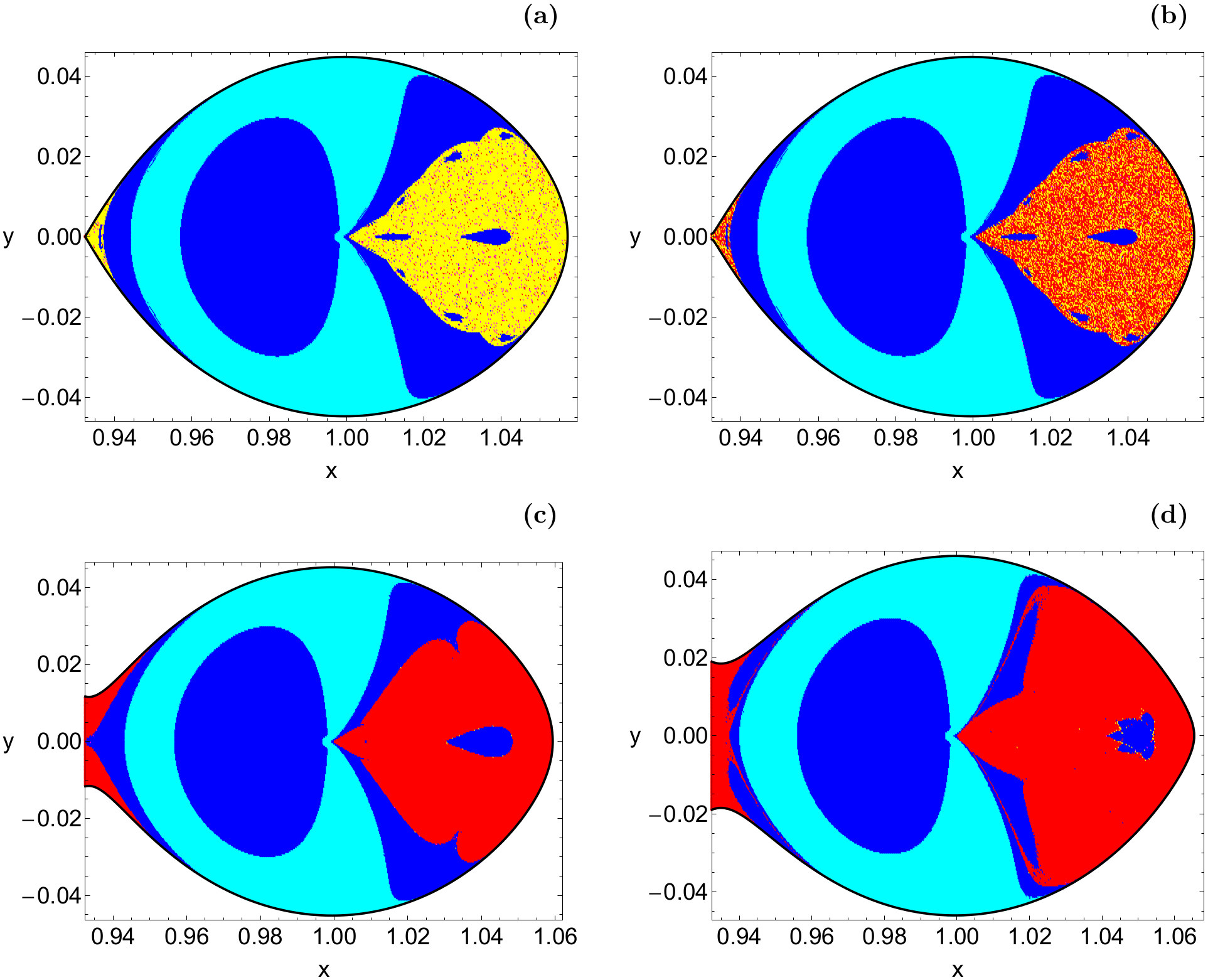}}
\caption{Colour-coded basin diagrams for Case II. (a-upper left): $C = 1.51933$, (b-upper right): $C = 1.519329$, (c-lower left): $C = 1.5191$, (d-lower right): $C = C_2$. The colour code is as follows: non-escaping regular orbits (blue), trapped sticky orbits (magenta), trapped chaotic orbits (yellow), collision orbits (cyan), escaping orbits to Sun realm (red). (Colour online only.)}
\label{hr2}
\end{figure*}

\begin{figure*}[!t]
\centering
\resizebox{\hsize}{!}{\includegraphics{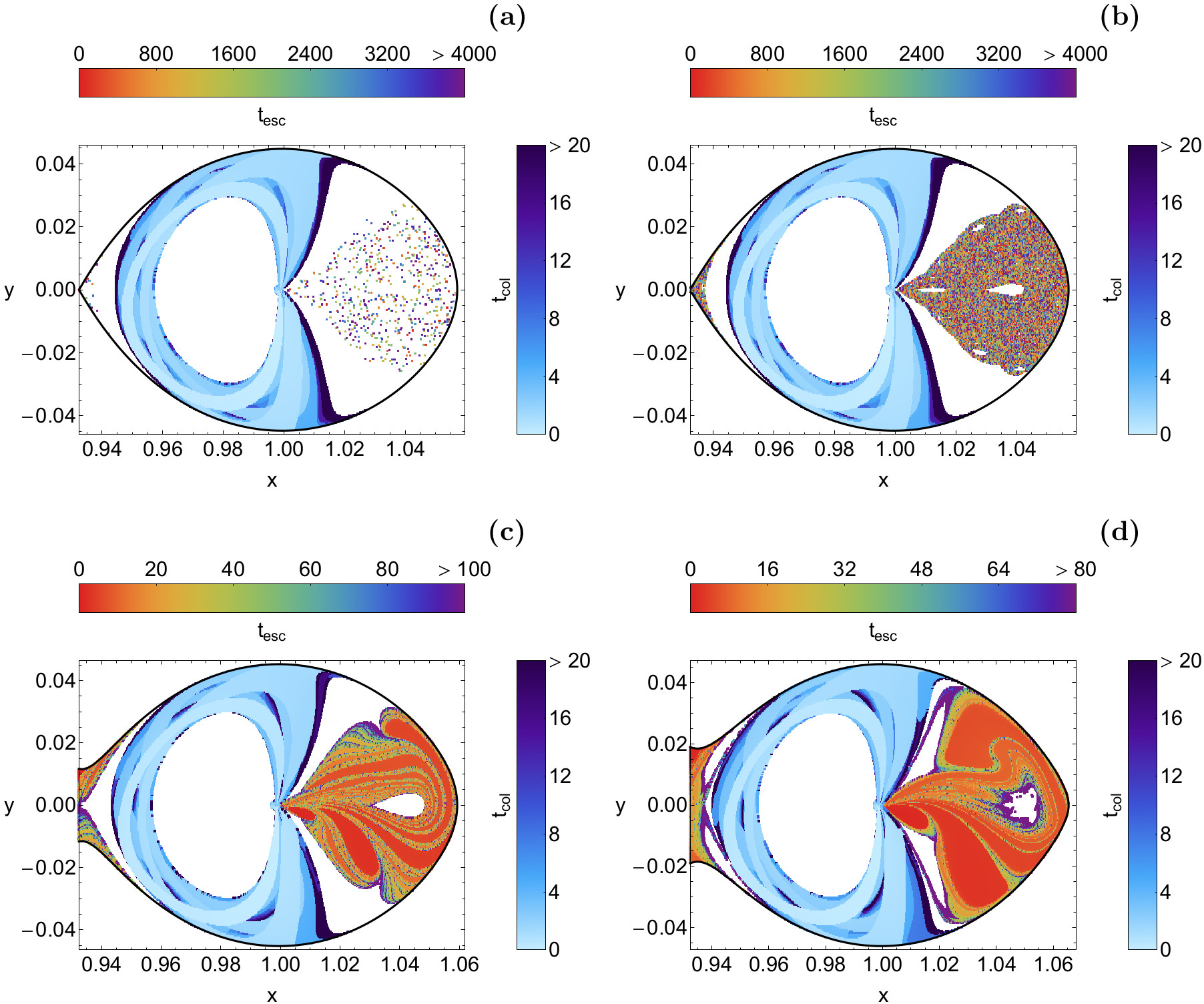}}
\caption{The corresponding distribution of the collision and escape time of the orbits for the values of the Jacobi constant of Fig. \ref{hr2}(a-d). The darker the colour, the larger the collision/escape time. Initial conditions of all types of bounded orbits are shown in white. (Colour online only.)}
\label{hr2t}
\end{figure*}

The next case considers the second Hill region configurations in which the test particle is allowed to move between Sun and Jupiter, through the open channel around $L_1$. The orbital structure of the configuration $(x,y)$ space is unveiled in Fig. \ref{hr2}(a-d), using colour-coded diagrams.

Our numerical calculations indicate that in the interval $1.51933 < C < C_1$, despite the existence of a bottleneck channel around $L_1$ (see Fig. \ref{isos}), there is no evidence of escaping orbits to the Sun domain. This clearly means that, for this range of the Jacobi constant values, the channel is very narrow and therefore all chaotic orbits need much more than 5000 dtu to eventually escape from the scattering region. In panel (a) of Fig. \ref{hr2}, where $C = 1.51933$, we see that inside the chaotic domain there is a fractal-like \cite{AVS01} mixture of sticky, chaotic and escaping orbits. At the boundaries of the chaotic region and along the symmetry axis, we notice a chain of six small stability islands (archipelago), mainly composed by quasi-periodic orbits. It should be noted that except for the existence of the stability archipelago at the frontier of the chaotic zone, the structure of the basin diagram is almost identical to the one presented earlier in panel (b) of Fig. \ref{hr1}. When $C = 1.519329$ it is seen in panel (b) of Fig. \ref{hr2} that the amount of escaping orbits increases rapidly, while for $C = 1.5191$ (see panel (c) of Fig. \ref{hr2}) all initial conditions of chaotic orbits disappear, thus giving place to escaping orbits to the Sun realm. At the same time, the secondary resonant orbits, that have been observed for $C = 1.519329$, are no longer present. When $C = C_2$, one may observe in panel (d) of Fig. \ref{hr2}, that other types of secondary resonant quasi-periodic orbits appear, while again about one-third of the configuration $(x,y)$ plane is occupied by initial conditions of escaping orbits. From the shapes of several types of basins, shown in Fig. \ref{hr2}(a-b), it becomes more evident that the parametric evolution of the bounded, collision as well as escape basins in the Sun-Jupiter system, is very different, compared to previous studies regarding celestial bodies in the Solar System \cite{Z15b}.

It deserves mentioning that in our numerical calculations we have followed the approach used in Ref. \refcite{dAT14} and of course in Refs. \refcite{Z15a,Z15b}, i.e., we consider that an orbit escapes to the Sun domain if the test particle passes through $L_1$, even if its final state is different for very long integration time. This assumption is supported by the fact that just a very small portion of orbits that initially enter the Sun realm return inside the scattering region.

In Fig. \ref{hr2t}(a-d), using tones of blue for collision orbits and a rainbow pallet for escaping orbits, we display the corresponding distributions for collision and escape times of the orbits. Once more, light colors correspond to short collision/esacpe times, dark colors indicate large collision/escape times, while white colour denotes all types of bounded motion. By inspecting the spatial distribution of the various different ranges of escape times, we are able to associate medium escape time with the stable manifold of a non-attracting chaotic invariant set, which spread out throughout the chaotic sea. On the other hand, the largest escape times are associated to the sticky orbits, surrounding the stability islands.

\subsection{Case III: $C_2 > C > C_3$}
\label{ec3}

\begin{figure*}[!t]
\centering
\resizebox{0.9\hsize}{!}{\includegraphics{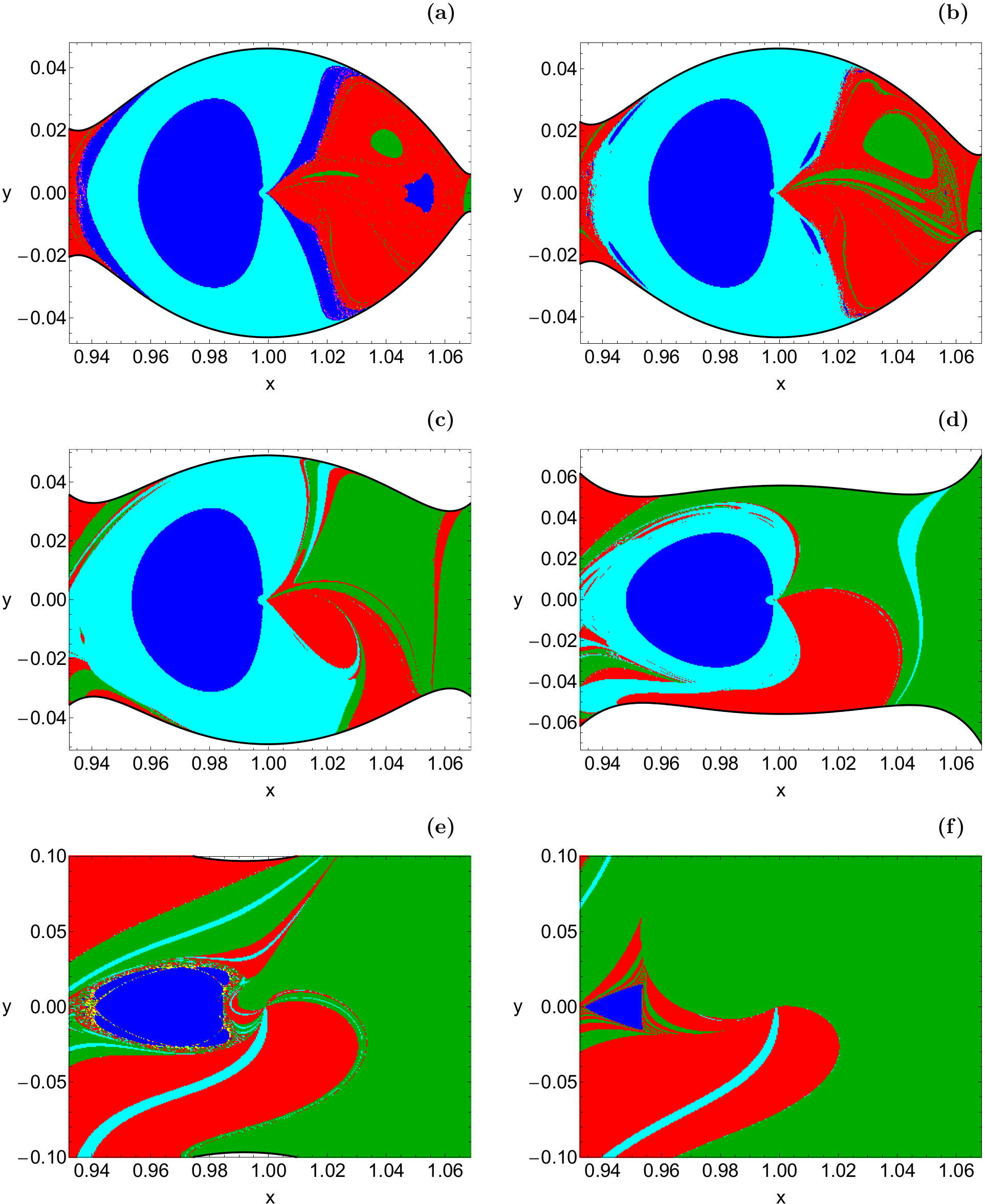}}
\caption{Colour-coded basin diagrams for Case III. (a): $C = 1.51865$, (b): $C = 1.5185$, (c): $C = 1.5175$, (d): $C = 1.5151$, (e): $C = 1.5079$, (f): $C = C_3$. The colour code is as follows: non-escaping regular orbits (blue), trapped sticky orbits (magenta), trapped chaotic orbits (yellow), collision orbits (cyan), escaping orbits to Sun realm (red), escaping orbits to the exterior realm (green). (Colour online only.)}
\label{hr3}
\end{figure*}

\begin{figure*}[!t]
\centering
\resizebox{0.85\hsize}{!}{\includegraphics{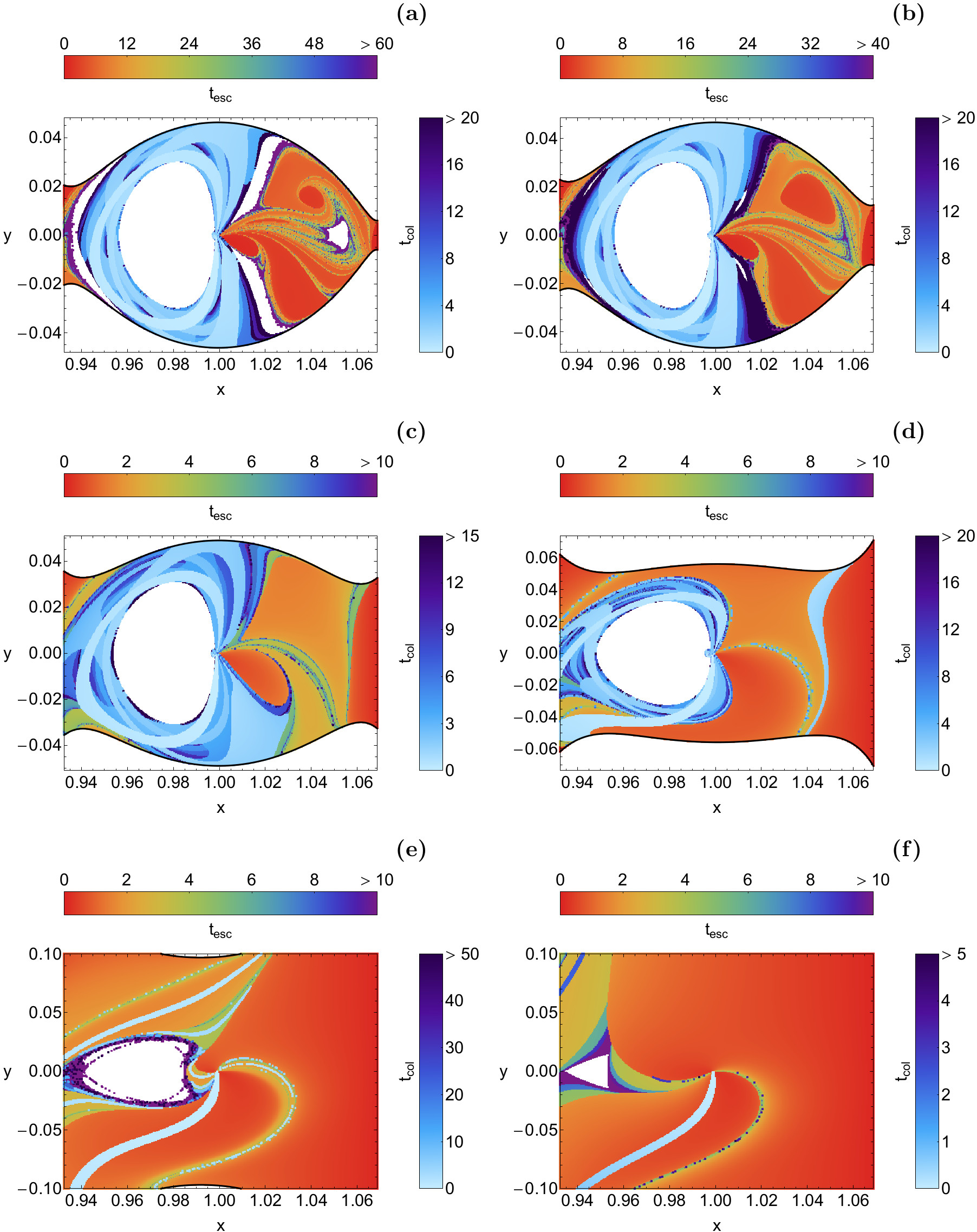}}
\caption{The corresponding distribution of the collision and escape time of the orbits for the values of the Jacobi constant of Fig. \ref{hr3}(a-f). The colour-code is the same as in Fig. \ref{hr2t}. (Colour online only.)}
\label{hr3t}
\end{figure*}

\begin{figure*}[!t]
\centering
\resizebox{\hsize}{!}{\includegraphics{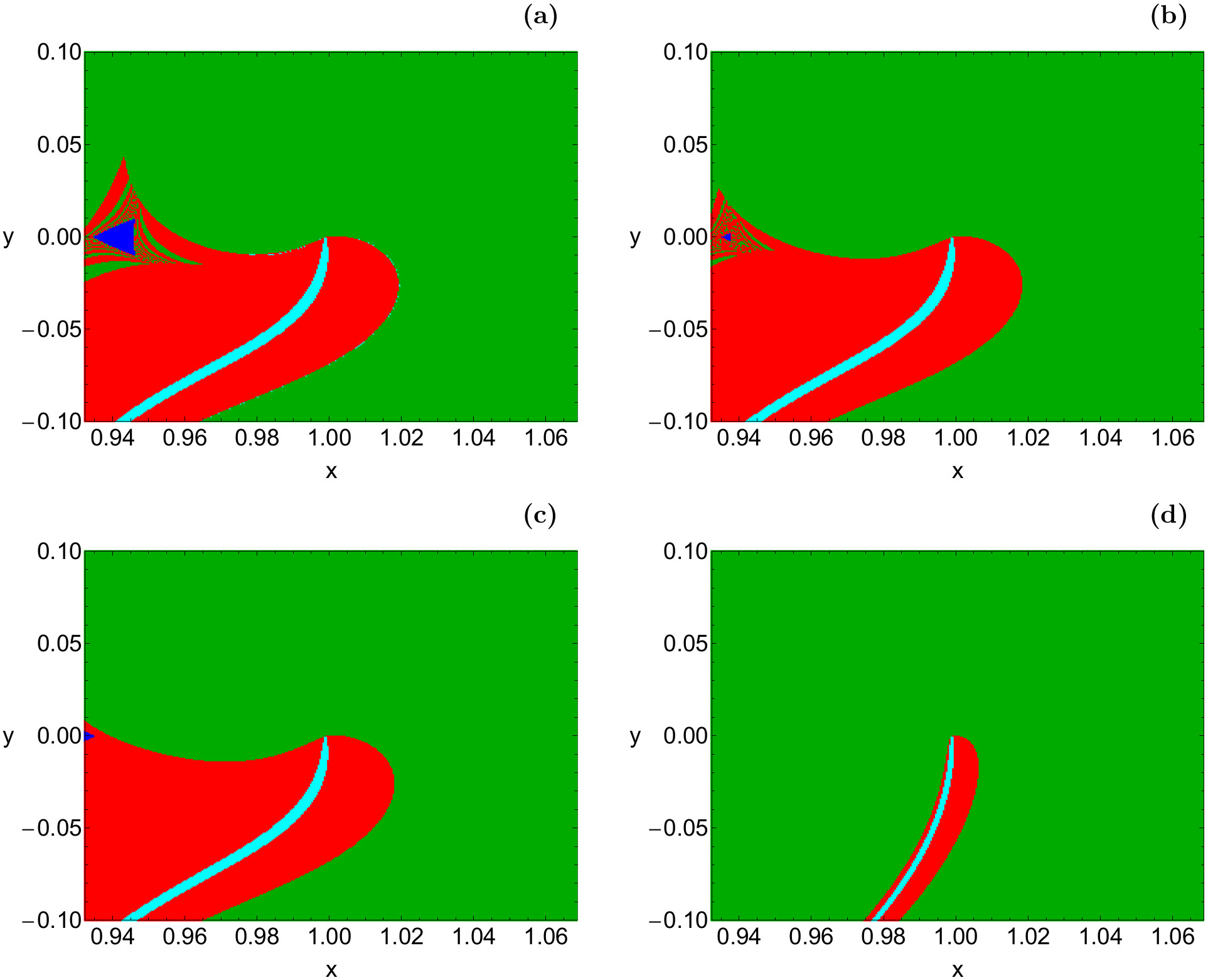}}
\caption{Colour-coded basin diagrams for Case IV. (a-upper left): $C = C_4$, (b-upper right): $C = 1.4985$, (c-lower left): $C = 1.4978$, (d-lower right): $C = 1.4$. The colour code is the same as in Fig. \ref{hr3}. (Colour online only.)}
\label{hr4}
\end{figure*}

\begin{figure*}[!t]
\centering
\resizebox{\hsize}{!}{\includegraphics{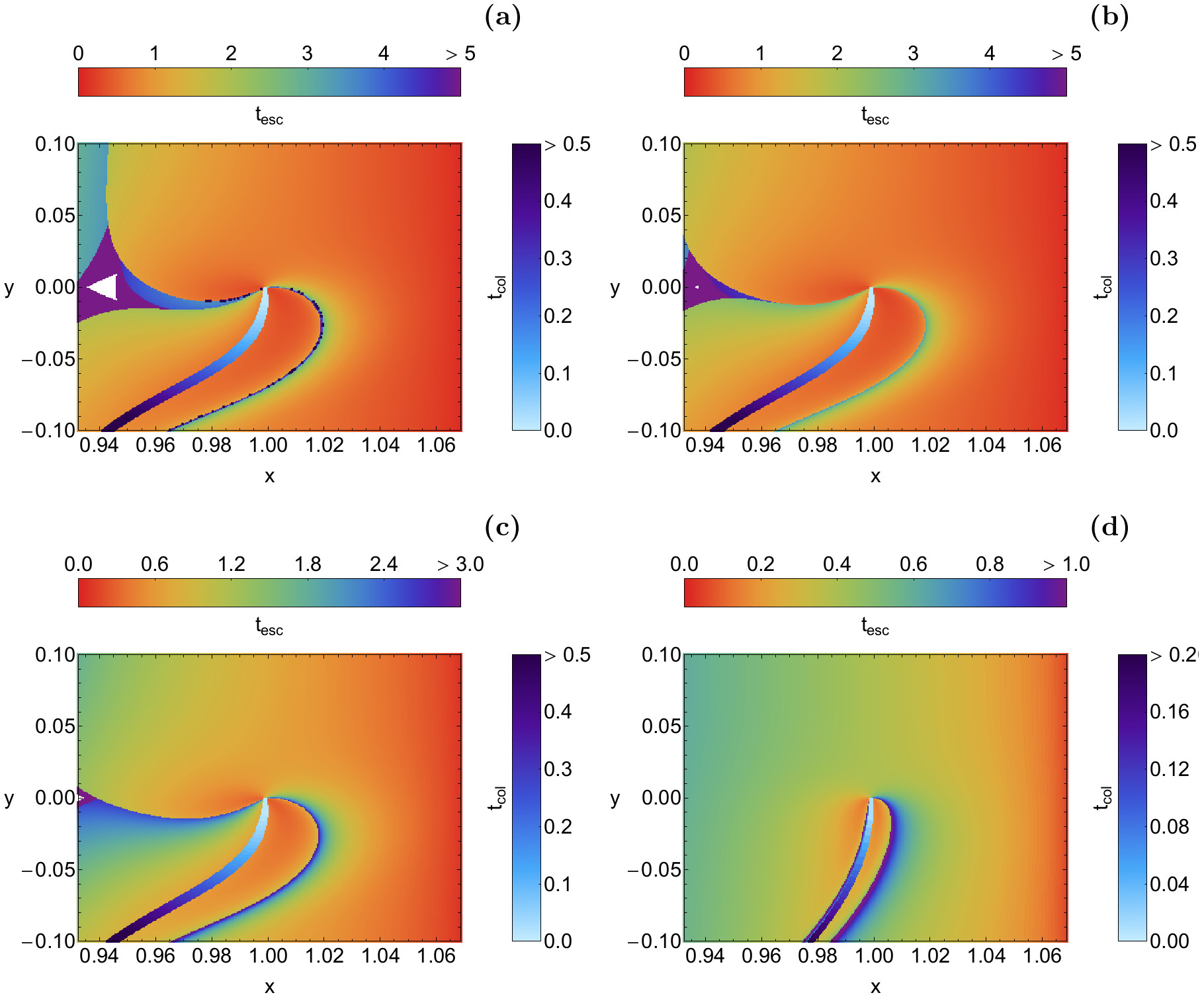}}
\caption{The corresponding distribution of escape and collision times for the values of the Jacobi constant of Fig. \ref{hr4}(a-d). The colour-code is the same as in Fig. \ref{hr2t}. (Colour online only.)}
\label{hr4t}
\end{figure*}

From the point of view of planetary systems and celestial mechanics, the case $C_2 > C > C_3$ constitutes the Hill region configurations with the most dynamical interest (the reader can find a detailed discussion about this topic in Ref. \refcite{dAT14}). When $C < C_2$ a second channel opens, around the Lagrange point $L_2$, thus allowing the test particle to enter the exterior region and escape from the system.

For $C = 1.51865$, we observe in panel (a) of Fig. \ref{hr3}, that several small basins of escape, corresponding to the exterior realm (green areas), emerge mainly at the right side of Jupiter, inside the Sun realm basin (red region). Note that by setting lower values of the Jacobi constant (i.e., higher values of the total orbital energy), the amount of orbits escaping to the exterior realm grows, while at the same time the rate of escaping orbits to the Sun realm, as well as that of collision orbits decreases. In panel (d) of Fig. \ref{hr3}, it can be seen that for $C = 1.5151$ the total percentage of escaping orbits occupies more that half (about 60\%) of the entire configuration $(x,y)$ plane. However, the portion of escaping orbits to the exterior realm is almost double with respect to that of the escaping orbits to the Sun realm. When $C = 1.5079$ (see (e) of Fig. \ref{hr3}) two important phenomena occur: (i) the main collision basin around the stability island, at the left side of Jupiter, disappears, while four small islands of secondary resonance emerge around the main stability island of 1:1 retrograde quasi-periodic orbits. (ii) A highly fractal mixture of all possible types of initial conditions (even trapped chaotic ones) surrounds the vicinity of the boundaries of the stability islands of non-escaping regular motion. If the Jacobi constant is equal to the critical value $C_3$, one may observe in panel (f) of Fig. \ref{hr3} that the initial conditions of orbits that escape to the exterior region dominate the configuration plane by occupying more than 70\% of its area. Furthermore, there is a weak presence of collision and bounded regular orbits around the Jupiter, which are manifested through the existence of small basins.

In Fig. \ref{hr3t}(a-f) we illustrate the corresponding distribution of the escape and collision time of orbits on the configuration $(x,y)$ space. A closer look at the time scale, given in the appended colour-bar, shows that in most of the cases, more than 90\% of the initial conditions of the orbits escape from the scattering region, in less than 10 dtu.

When $C_3 > C > C_4$ the channel around $L_3$ opens, thus allowing orbits to escape to the exterior region also from the left side of the primary (Sun). However, since we decided to focus our study in the vicinity of the secondary (Jupiter), this escape channel in the ZVC is not visible and therefore this energy case has limited physical meaning in our study.

\subsection{Case IV: $C \leq C_4$}
\label{ec4}

The last case under consideration involves the scenario when the test particle can freely travel all over the configuration $(x,y)$ plane with no restrictions of energetically forbidden regions. Again, all the different aspects of the numerical approach remain exactly the same as in the previously studied cases.

Fig. \ref{hr4}(a-d) reveals the parametric evolution of the orbital structure of the configuration space, through the colour-coded diagrams. The most important phenomena which take place, as the value of the Jacobi constant decreases, are the following:
\begin{itemize}
  \item The area on the $(x,y)$ plane covered by initial conditions of orbits that escape to the exterior region grows rapidly and for $C < 1.4$ they occupy more than 95\% of the same plane.
  \item The area of the basin composed of initial conditions of orbits that escape to the Sun realm constantly decreases.
  \item The portion of the initial conditions that lead to collision with Jupiter seems to be almost unperturbed by the change on the value of the Jacobi constant. In particular, a thin collision basin is always present, which seems to survive even at very low values of $C$, or equivalently, at extremely high levels of the total orbital energy.
  \item We find no numerical evidence of non-escaping regular orbits for $C < 1.4978$, since the stability islands disappear. Therefore, we may infer that bounded motion is not possible at this range of the Jacobi values.
\end{itemize}

The corresponding distribution of the collision and escape time of orbits on the configuration space is depicted in Fig. \ref{hr4t}(a-d). Our results suggest that the average escape time of the orbits decreases for lower values of the Jacobi constant. Moreover, at this energy case $(C \leq C_4)$ the collision to the secondary is very fast, because the vast majority of the corresponding initial conditions of the orbits have collision rates lower than about 0.5 dtu.

\subsection{A summarized analysis of the numerical results}
\label{over}

\begin{figure}[!t]
\begin{center}
\includegraphics[width=0.6\hsize]{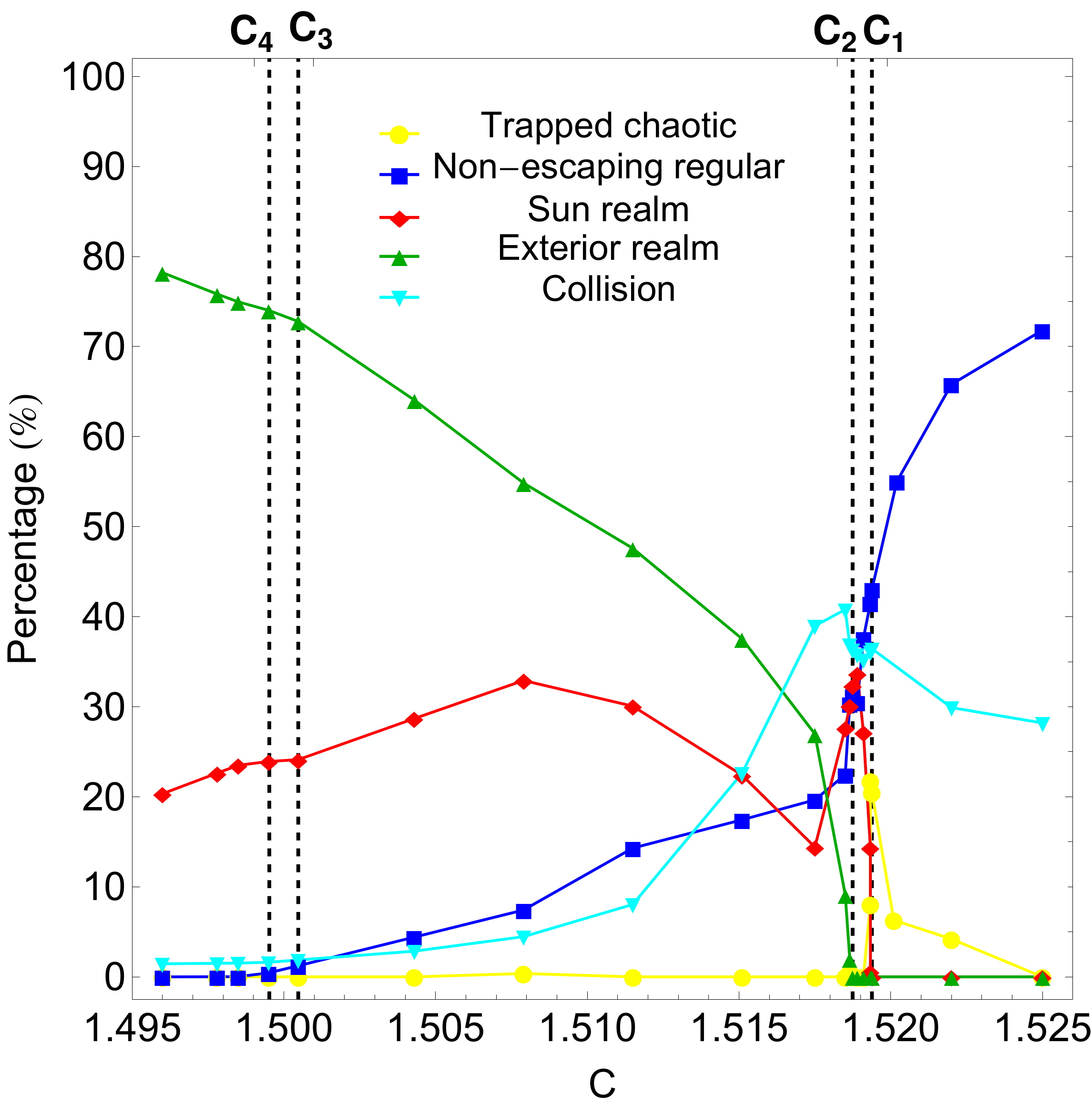}
\end{center}
\caption{Parametric evolution of the percentages of all types of orbits with initial conditions on the configuration $(x,y)$ plane, in terms of the Jacobi constant $C$. The vertical dashed black lines indicate the four critical values of $C$. (Colour online only.)}
\label{percs}
\end{figure}

\begin{figure*}[!t]
\centering
\resizebox{\hsize}{!}{\includegraphics{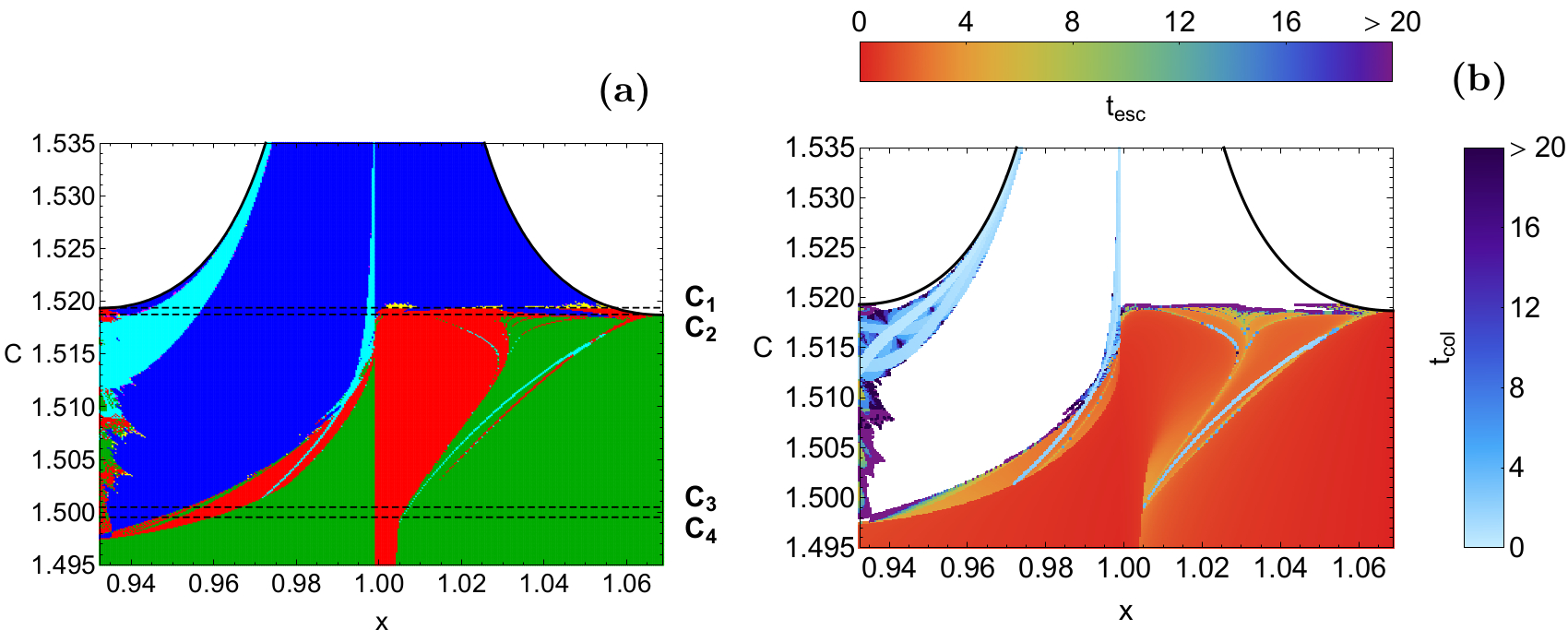}}
\caption{(a-left): Orbital structure of the $(x,C)$ plane. (b-right): The distribution of the corresponding collision and escape time of the orbits. The colour codes are the same as in Figs. \ref{hr3} and \ref{hr3t}, respectively. (Colour online only.)}
\label{xct}
\end{figure*}

\begin{figure*}[!t]
\centering
\resizebox{\hsize}{!}{\includegraphics{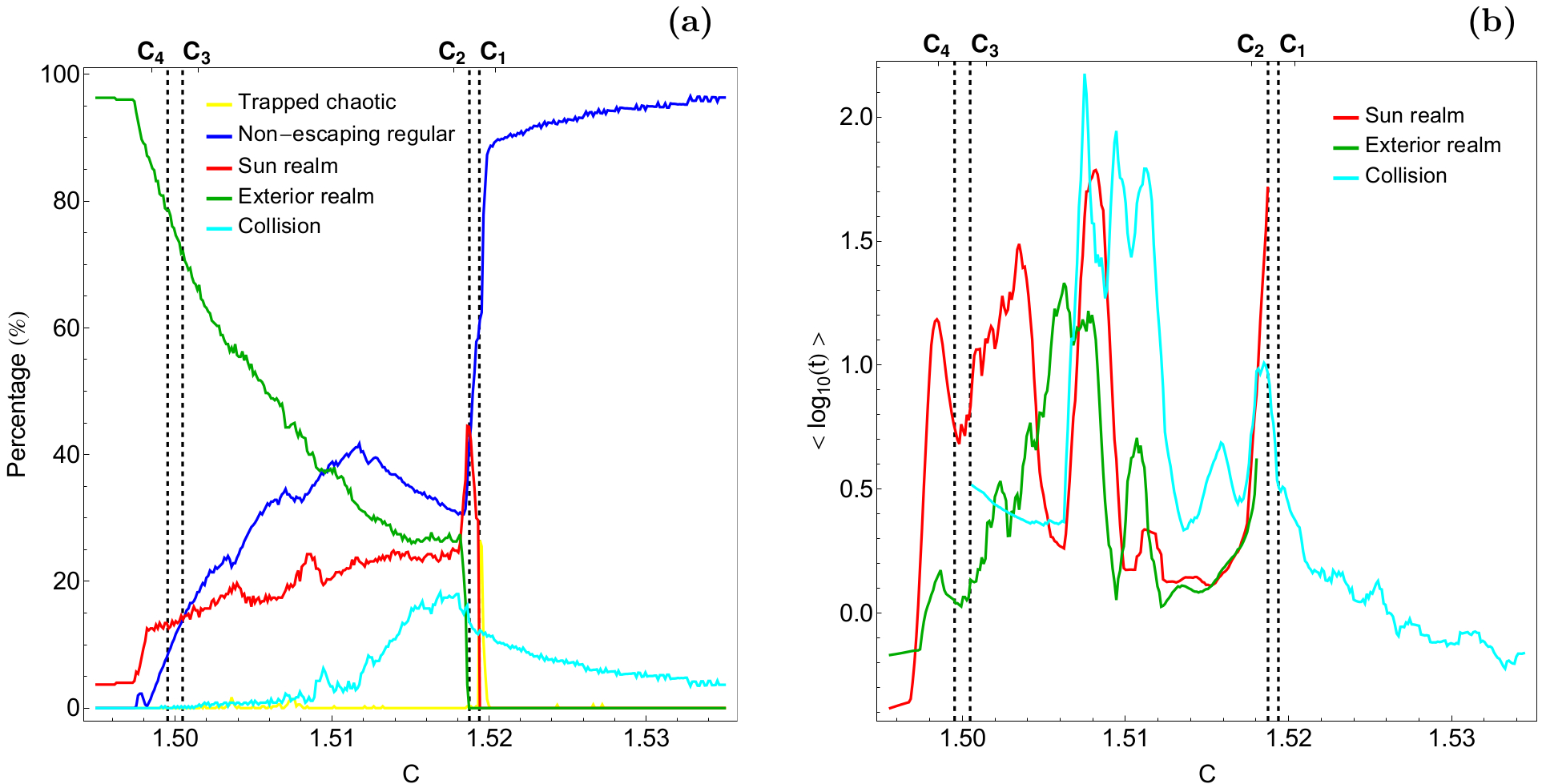}}
\caption{Evolution of (a-left): the percentages of all types of orbits and (b-right): the average logarithmic collision and escape time $(< \log_{10}(t) >)$ of orbits on the $(x,C)$-plane, as a function of the value of the Jacobi constant $C$. (Colour online only.)}
\label{stats}
\end{figure*}

It would be very informative for the reader to observe a summarized monitoring of the percentage of orbits pertaining to the different categories, introduced in section \ref{cometh}. In Fig. \ref{percs}, we show a diagram with the parametric evolution of all types of percentages, as a function of the Jacobi constant $C$. From this figure, we can see that at very high values of $C$, regular bounded motion is the most populated type of motion, occupying more than 70\% of the configuration space. However, as the value of the Jacobi constant decreases the percentage of non-escaping regular orbits diminishes until $C = C_2$, where at this point, the magnitude of the slope changes and the percentage reduction continues much slower. Bounded regular motion is possible up to about $C = 1.4975$, while for smaller values of $C$ there is no numerical evidence of this kind of motion. For most of the investigated range of the Jacobi values chaotic motion corresponds to extremely low percentages (less than 0.5\%), the only observed peak occurs at exactly $C = C_1$ and it is equal to about 22\%. As the channel around $L_1$ opens for $C < C_1$ the percentage of escaping orbits to the Sun realm increases until $C = C_2$, while for lower values of $C$ the trend is reversed. For $C < C_2$ the second channel around the Lagrange point $L_2$ also opens and the percentage of orbits escaping to the exterior realm displays a rapid increase. For $C < 1.5164$ escaping orbits to the exterior realm dominate the $(x,y)$ plane, while for extremely low values of the Jacobi constant $(C < C_4)$ they occupy more than 90\% of the configuration space. The rate of collision orbits gets reduced as soon as $C < C_2$ and tends asymptotically to zero. However, our analysis indicates that collision motion is possible for all tested values of $C$, even at extremely low percentages (less than 1\%).

Taking into account the above-mentioned analysis, we may say that the evolution of the percentages of each basin in the Sun-Jupiter system is very similar to that observed for the Pluto-Charon system in Ref. \refcite{Z15b}. The main difference concerns the interval $C < C_4$, where for the Pluto-Charon system we had used more complicated escape criteria (more escape sectors), thus leading to different percentages.

The colour-coded diagrams in the configuration $(x,y)$ space provide enough information about the phase space mixing. However, such analysis is performed only for fixed values of the Jacobi constant and for orbits that traverse the surface of section, either progradely or retrogradely. In order to surmount these barriers, Ref. \refcite{H69}, introduced a new plane of representation that can supply useful information about the classification of the orbits by using the section $y = \dot{x} = 0$, $\dot{y} > 0$. With this new approach the Jacobi integral can be used as an independent variable, and therefore, the orbital structure of the Sun-Jupiter system can be monitored using a continuous spectrum of values of $C$. In panel (a) of Fig. \ref{xct} we present the orbital structure of the $(x,C)$ plane when $x \in [x(L_1),x(L_2)]$ and $C \in [1.495,1.535]$, while in panel (b) of the same figure we display the distributions of the corresponding collision and escape time of the orbits. The frontier between energetically allowed and not-allowed motion is depicted with a black solid line which is defined as
\begin{equation}
f(x,C) = J(x,y=0,\dot{x}=0,\dot{y}=0) = C,
\label{zvc}
\end{equation}
while the critical values of $C$ are pointed by horizontal dashed black lines.

The two stability islands, corresponding to prograde and retrograde motion around Jupiter are now visible. Once more it is manifest that choosing smaller values of the Jacobi constant $C$, induces a larger number of orbits escaping to the exterior domain through $L_2$, while larger values of $C$ lead to an overpopulation of bounded orbits. Just below the right stability island, which ends at about $C = C_2$, there is a fractal mixture of initial conditions corresponding to both types of escaping orbits. The colour-coded diagram shows us exactly how the fractality of the several basin boundaries strongly depends not only on the Jacobi constant but also on the spatial variable. In particular, one can observe a very interesting phenomenon. It is seen that the fractality of the basin boundaries, which is related to the unpredictability, migrates from the upper right side to the lower left side of the secondary (Jupiter), for low values of the Jacobi constant (i.e. high values of the total orbital energy).

The parametric evolution of the percentages of all types of orbits on the $(x,C)$ plane is (see panel (a) of Fig. \ref{stats}), in general terms, very similar to that discussed earlier in Fig. \ref{percs}, for the configuration $(x,y)$ plane. The combined analysis of the orbit classification, in both types of planes, strongly suggests that at very high values of the Jacobi constant $C$ bounded regular motion completely dominate all types of planes. At very low values of $C$ on the other hand, escaping motion to the exterior realm is, by far, the mots populated type of motion. It would be also interesting to shed some light to the collision as the well as to the escape time of the orbits. The evolution of the average logarithmic value of the collision and escape time $(< \log_{10}(t) >)$ of the orbits on the $(x,C)$ plane, as a function of the value of the Jacobi constant $C$, is given in panel (b) of Fig. \ref{stats}. It is seen there that in most of the cases the average escape time of the orbits to the exterior realm is lower than the average escape time of the orbits to the Sun realm. It is interesting to note that the peak of the escape time of the Sun realm is about 63 time units, when $C = 1.508$, while the peak of the escape time of the exterior realm is only about 20 time units, when $C = 1.506$. The peak of the collision time is observed when $C = 1.5075$ and it is equal to about 158 time units.

\begin{figure}[!t]
\begin{center}
\includegraphics[width=0.6\hsize]{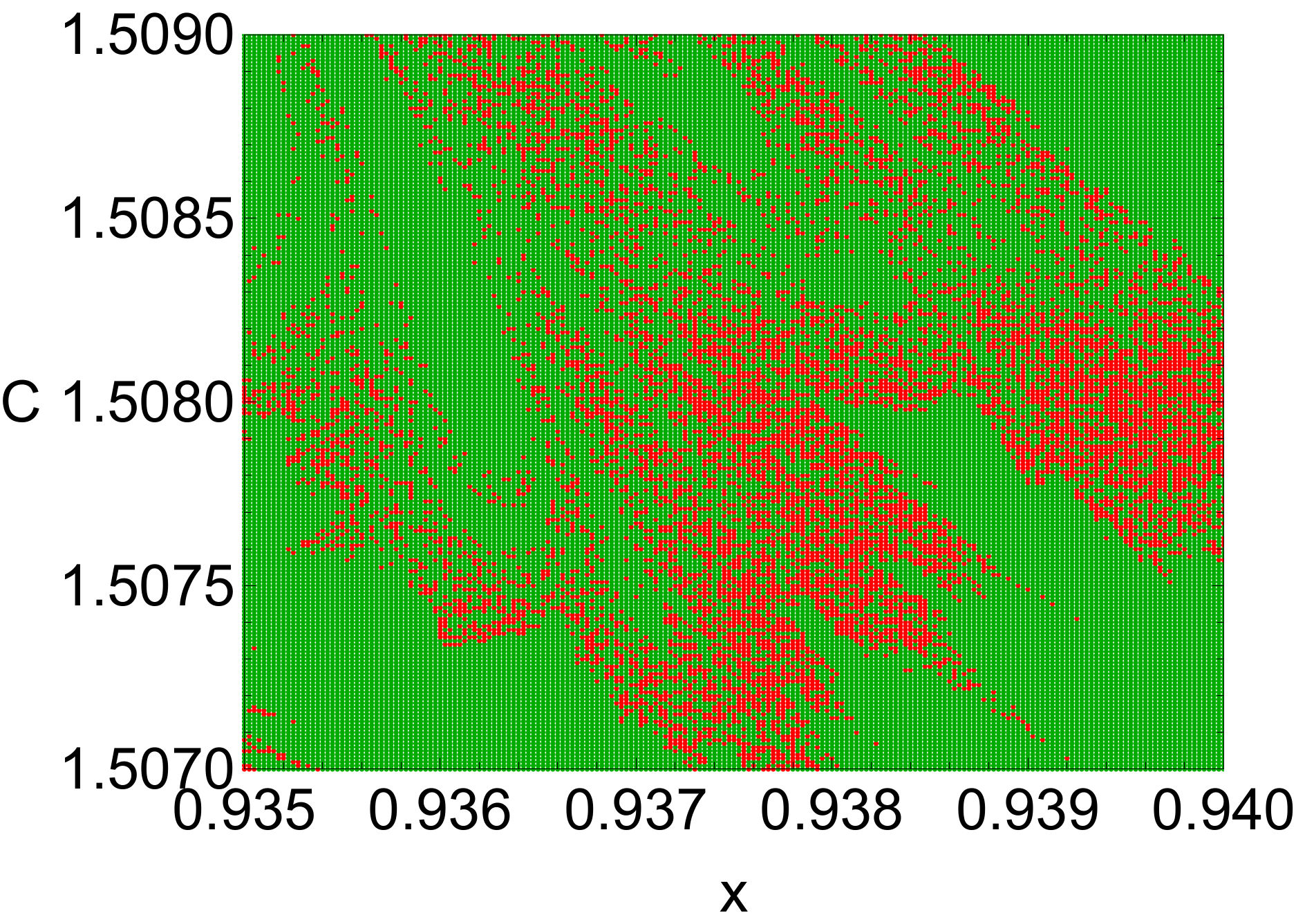}
\end{center}
\caption{Orbital structure of a local region on the $(x,C)$ plane. Initial conditions with the same final state in both classical Newtonian and PN dynamics are shown in green, while converging initial conditions, with different final states, are shown in red. (Colour online only.)}
\label{dif}
\end{figure}

Before ending this section, we would like to analyse the differences on the orbital structure between the classical Newtonian (CN) dynamics $(\epsilon = 0)$ and the PN dynamics $(\epsilon = 1)$. Our numerical analysis strongly suggests that in general terms the influence of the PN terms is rather weak, however it does affect the final state of the orbits. To prove this we reclassified the initial conditions of the orbits on the $(x,C)$ plane (see Fig. \ref{xct}), by setting $\epsilon = 0$. The corresponding colour-coded diagram, with the classification of the orbits, looks almost identical to that shown in panel (a) of Fig. \ref{xct}. More precisely, only about 12\% of the total initial conditions displays a different final state between CN and PN. In order to observe the differences on the final state of the orbits, due to the inclusion or not of the PN terms, we have to zoom in and examine local regions of the $(x,C)$ plane. In Fig. \ref{dif} we see a local region of the $(x,C)$ plane, with the highest portion of converging initial conditions. Green colour indicated initial conditions with the same final state in both CN and PN, while the divergent initial conditions are shown in red. It is seen that both types of initial conditions of orbits are not randomly distributed on the $(x,C)$ plane, as we can distinguish basins of converging and not converging initial conditions. Additional computations reveal that inside the local region, shown in Fig. \ref{dif}, all possible divergent initial conditions exist. In particular, since there are five main final states (trapped chaotic, non-escaping regular, collision, escaping to Sun realm and escaping to the exterior region) and two different dynamical cases (CN and PN) the corresponding total number of permutations is $N = 5!/(5-2)! = 20$. In Table \ref{tab2} we provide the exact initial conditions of 20 characteristic examples corresponding to all possible divergent final states between CN and PN.

\begin{table}
 \center
 \caption{Divergent initial condition $(x_0,C_0)$ between the two dynamical cases: classical Newtonian dynamics (CN) and post-Newtonian dynamics (PN). The numerical code denoting the final states is the following: trapped chaotic orbit (-1); non-escaping regular orbit (0); escaping to Sun realm (1); escaping to the exterior region (2); collision orbit (9).}
 \label{tab2}
 \begin{tabular}{ccrr}
  \hline
  $C_0$ & $x_0$ & CN & PN \\
  \hline\hline
  1.50708 & 0.939075 &  0 & -1 \\
  1.50701 & 0.935000 & -1 &  0 \\
  1.50702 & 0.935250 &  0 &  1 \\
  1.50702 & 0.937350 &  0 &  2 \\
  1.50705 & 0.935100 &  1 &  0 \\
  1.50703 & 0.937355 &  2 &  0 \\
  1.50708 & 0.937300 &  0 &  9 \\
  1.50709 & 0.937350 &  9 &  0 \\
  1.50811 & 0.938225 &  1 &  2 \\
  1.50804 & 0.938450 &  2 &  1 \\
  1.50768 & 0.937500 &  1 &  9 \\
  1.50811 & 0.939150 &  2 &  9 \\
  1.50700 & 0.937750 &  9 &  1 \\
  1.50700 & 0.935375 &  9 &  2 \\
  1.50839 & 0.936675 &  1 & -1 \\
  1.50709 & 0.937775 &  2 & -1 \\
  1.50881 & 0.938725 & -1 &  1 \\
  1.50869 & 0.939200 & -1 &  2 \\
  1.50711 & 0.937400 &  9 & -1 \\
  1.50717 & 0.937800 & -1 &  9 \\
  \hline
 \end{tabular}
\end{table}

\section{Discussion}
\label{disc}

The orbital dynamics of a small body (e.g., a spacecraft, comet, meteor or asteroid) in the presence of the Sun-Jupiter system has been numerically investigated. By using the PN equations of motion for the planar circular restricted three-body problem, we have performed an orbital classification in a scattering region around Jupiter. To do so, we have determined the position of the Lagrange points and the corresponding values of the Jacobi constant at these points. After numerically integrating several sets of initial conditions, for all possible Hill region configurations, we managed to classify the orbits into four main categories: bounded orbits around Jupiter, escaping orbits to Sun realm, escaping orbits to the exterior realm and orbits that lead to collisions with Jupiter. Furthermore, the SALI chaos indicator has been used in order to further classify bounded orbits into three sub-categories: regular orbits, sticky orbits, and chaotic orbits.

Despite the fact that the PN correction terms are modulated by a factor of $10^{-8}$, it is clear from our results that, in general, the differences on the Newtonian and PN Lagrange points are non-negligible, mainly for the triangular points $L_4$ and $L_5$ (approx. 1 km). Our result is in agreement with previous findings for the post-Newtonian collinear points reported in Ref. \refcite{YA12}, where the authors have found that the correction of distance for the triangular points is 923 m. Consequently, the final states of the orbits in both approaches can be significantly different. Our numerical analysis strongly suggests that a refined version of the Sun-Jupiter system, that includes general relativistic correction terms, could certainly update most of our current knowledge about the dynamics of this particular planetary system.

It is important to note that along the paper we neglect the perturbations produced by other planets. This approximation is based on the fact that despite the relativistic contribution, due to the Sun, is 6 orders of magnitude smaller than the one of the Newtonian terms, this contribution is 4 orders of magnitude larger than the effect due to the presence of the second most-massive planet in the Solar System, i.e. Saturn.

As far as we know, this is the first time that the nature of motion, in the vicinity of Jupiter, is explored in such a thorough and systematic manner, through the orbit classification using the two-dimensional colour coded diagrams. Therefore we may claim that our paper adds considerable new information to the field of planetary and celestial dynamics.

We hope that the current numerical results to be useful in the active field of the orbital dynamics of the PN version of the restricted three-body problem. Taking into account that our present outcomes are positive, as well as encouraging, it is in our future plans to expand our investigation into three dimensions, thus revealing the orbital content of the full six-dimensional phase space.

\section*{Acknowledgments}

One of the authors (FLD) gratefully acknowledges the financial support provided by Universidad de los Llanos and COLCIENCIAS, Colombia, under Grant No. 8840. We would like to express our warmest thanks to the anonymous referee for the careful reading of the manuscript and for all the apt suggestions and comments which allowed us to improve both the quality and the clarity of our paper.

\appendix

\section{Relationship between physical and dimensionless units}
\label{Ap1}

The planar circular restricted three body problem has three natural scales associated to the three fundamental mechanical quantities (time, distance, and mass): the distance between the primaries $a$, the total mass of the system $M = m_{1} + m_{2}$, and the angular velocity of their orbital motion $\omega$. In the canonical system of units, introduced in section \ref{mod} all this three quantities are dimensionless and normalized to 1, i.e. $a = M = 1$ and $\omega \approx 1$. Therefore it implies that in the resulting system of units the kinematic quantities are also dimensionless.

Taking into account that the average distance between Jupiter and the Sun is 5.20336301 AU, and the orbital period of Jupiter is about 11.8701 years, the conversion of astronomical units and years to the canonical units are as follows: 1 AU = 1/5.20336301 and 1 year = $2\pi/11.8701$. On the other hand, the speed of light in astronomical units per year reads as $c = 63232.78$AU/yr. Then in our system of units the speed of light is equal to $c = 22945.23619$, for the case of Sun-Jupiter \cite{RPKV01,LC15}.

As can be noted, we have assumed that the post-Newtonian contribution to the angular velocity is too small, i.e $\omega = 1 + \omega_{1}/c^2 \approx 1$. In order to prove this assumption, let us start by considering that unlike the classical Newtonian system, the post-Newtonian angular velocity depends on the mass parameter $\mu$ and the speed of light $c$, i.e.
\begin{equation}
\omega = 1 + \frac{\mu(1 - \mu) - 3}{2 c^2},
\label{om1}
\end{equation}
where the mass parameter, for the Sun-Jupiter system, is given by $\mu = 0.000953817733371$, and the velocities ratio is $v/c = 0.000043582$. Moreover $v = \omega a$, such that $\omega/c = v/c$ and from Eq. \eqref{om1}
\begin{equation}
\frac{v}{c} = \frac{1}{c}\left(1 + \frac{\mu(1 - \mu) - 3}{2 c^2}\right),
\label{velr}
\end{equation}
replacing the numerical values for $v/c$ and $\mu$, and solving for $c$ we get $c = 22945.23612$, which is practically the same value calculated with the assumption $\omega \approx 1$.

\section{Coefficients for the Lagrange points}
\label{Ap2}

The exact numerical coefficients entering equations \eqref{col} and \eqref{triag} are the following:

\begin{align*}
 C_{0}^{1}        &=  1.000000000000055; &C_{1}^{1}         &=-0.717348891565242; \\
 C_{2}^{1}        &=  0.272843046490294; &C_{3}^{1}         &= 0.254713513239888; \\
 C_{4}^{1}        &=  0.231448423015461; &C_{5}^{1}         &= 0.220945106759323; \\
 C_{6}^{1}        &=  0.141234754066445; &C_{7}^{1}         &= 0.395266691687848; \\
 C_{8}^{1}        &= -0.730153372527948; &C_{9}^{1}         &= 2.950558340203948; \\
 C_{10}^{1}       &= -7.029887711788199; &C_{11}^{1}        &=14.103048574726300; \\
 C_{12}^{1}       &=-20.298440525190710; &C_{13}^{1}        &=20.968507093895010; \\
 C_{14}^{1}       &=-13.495144532299240; &C_{15}^{1}        &= 4.392951206288244; \\
 \tilde{C}_{1}^{1}&= -0.693361274074247; &\tilde{C}_{2}^{1} &= 0.160249963115877; \\
 \tilde{C}_{3}^{1}&= -0.245614411667766; &\tilde{C}_{4}^{1} &= 0.065632595400727; \\
 \tilde{C}_{5}^{1}&= -0.099641922793147; &\tilde{C}_{6}^{1} &= 0.022606046410609; \\
 \tilde{C}_{7}^{1}&=  0.005895377094297; &\tilde{C}_{8}^{1} &= 0.047904193249469; \\
 \tilde{C}_{9}^{1}&= -0.040585826796281; &\tilde{C}_{10}^{1}&= 0.009028760630797;
\end{align*}

\begin{align*}
 C_{0}^{2}        &=  1.000000000044847; &C_{1}^{2}         &=-0.768734223717043; \\
 C_{2}^{2}        &=  0.210715408460650; &C_{3}^{2}         &= 0.230780785242519; \\
 C_{4}^{2}        &=  0.228323516648158; &C_{5}^{2}         &= 0.265414329835819; \\
 C_{6}^{2}        &=  0.086784389035090; &C_{7}^{2}         &= 0.784093293530970; \\
 C_{8}^{2}        &= -1.899462212900000; &C_{9}^{2}         &= 6.550074615819714; \\
 C_{10}^{2}       &=-15.400049404819560; &C_{11}^{2}        &=29.724426983783020; \\
 C_{12}^{2}       &=-41.885449085729490; &C_{13}^{2}        &=42.267441686705390; \\
 C_{14}^{2}       &=-26.737590667890560; &C_{15}^{2}        &= 8.505043708659647; \\
 \tilde{C}_{1}^{2}&=  0.693361270290015; &\tilde{C}_{2}^{2} &= 0.160250078262299; \\
 \tilde{C}_{3}^{2}&= -0.268305351636877; &\tilde{C}_{4}^{2} &=-0.088420841993041; \\
 \tilde{C}_{5}^{2}&= -0.078760868402773; &\tilde{C}_{6}^{2} &= 0.009924885128025; \\
 \tilde{C}_{7}^{2}&=  0.029056187821197; &\tilde{C}_{8}^{2} &= 0.060670725036652; \\
 \tilde{C}_{9}^{2}&= -0.058563087000968; &\tilde{C}_{10}^{2}&= 0.013578265351199;
\end{align*}

\begin{align*}
 C_{0}^{3}         &=-1.000000000000140;               &C_{1}^{3}        &=-0.310134696468904; \\
 C_{2}^{3}         &= 0.083520399808257;               &C_{3}^{3}        &= 0.118963335836620; \\
 C_{4}^{3}         &= 0.080317152559464;               &C_{5}^{3}        &= 0.057437704575241; \\
 C_{6}^{3}         &= 0.042753232135546;               &C_{7}^{3}        &= 0.028020399356123; \\
 C_{8}^{3}         &= 0.027537868347720;               &C_{9}^{3}        &= 0.001670004689334; \\
 C_{10}^{3}        &= 0.020696733440176;               &C_{11}^{3}       &= 0.0; C_{12}^{3}=0.0; C_{13}^{3}=0.0; \\
 C_{14}^{3}        &= 0.0; C_{15}^{3}=0.0;             &\tilde{C}_{1}^{3}&=-1.2629807526629\times10^{-10}; \\
 \tilde{C}_{2}^{3} &= 8.0000398616382\times 10^{-9}; &\tilde{C}_{3}^{3}&=-0.106532175762120; \\
 \tilde{C}_{4}^{3} &= 2.9725723912455\times 10^{-6}; &\tilde{C}_{5}^{3}&=-0.000026613361181; \\
 \tilde{C}_{6}^{3} &= 0.023168364513464;               &\tilde{C}_{7}^{3}&=-0.000619963259894; \\
 \tilde{C}_{8}^{3} &= 0.001626329250906;               &\tilde{C}_{9}^{3}&=-0.006770166905911; \\
 \tilde{C}_{10}^{3}&= 0.002229210191938;
\end{align*}

\begin{align*}
T_{0}^{4}        &= 0.500000001187120;                & T_{1}^{4}        &=-1.000000002374239; \\
T_{2}^{4}        &= 2.4722448217899\times 10^{-15}; & \tilde{T}_{0}^{4}&= 0.866025403099055; \\
\tilde{T}_{1}^{4}&= 8.2246306076289\times10^{-10};  & \tilde{T}_{2}^{4}&=-8.2246290911674\times10^{-10};
\end{align*}

\begin{align*}
T_{0}^{5}        &= 0.500000001187120;                & T_{1}^{5}        &=-1.000000002374239;\\
T_{2}^{5}        &= 2.4722448217899\times 10^{-15}; & \tilde{T}_{0}^{5}&=-0.866025403099055;\\
\tilde{T}_{1}^{5}&=-8.2246306076289\times10^{-10};  & \tilde{T}_{2}^{5}&= 8.2246290911674\times10^{-10};
\end{align*}

\end{document}